\documentclass{svjour3}                     % onecolumn (standard format)
\smartqed  % flush right qed marks, e.g. at end of proof
\usepackage{graphicx}
\usepackage{amsmath}
\usepackage{wasysym}
\begin{document}

\title{Behavior of nearby synchronous rotation of a Poincar\'e-Hough satellite at low eccentricity}

\titlerunning{Poincar\'e model and synchronous rotatiom}        % if too long for running head

\author{Beno\^it Noyelles}


\institute{FUNDP - The University of Namur \\
NAmur Center for Complex SYStems (NAXYS) \\
Belgium \\
\and IMCCE (Paris Observatory / USTL / UPMC) \\
France \\
\email{benoit.noyelles@fundp.ac.be}}

\date{Received: date / Accepted: date}

\maketitle

\begin{abstract}

 \par This paper presents a study of the Poincar\'e-Hough model of rotation of the synchronous natural satellites, in which these 
bodies are assumed to be composed of a rigid mantle and a triaxial cavity filled with inviscid fluid of constant uniform density and 
vorticity. In considering an Io-like body on a low eccentricity orbit, we describe the different possible behaviors of the system, depending 
on the size, polar flattening and shape of the core.

% \par We use for that the numerical tool. Starting from a Hamiltonian formulation of the system, we derive the Hamilton equations
%before integrating them numerically. Then, we use a frequency analysis algorithm to give a quasi-periodic representation, allowing
%us to split the different contributions and to characterise the equilibrium of the system.

\par We use for that the numerical tool. We propagate numerically the Hamilton equations of the systems, before expressing the resulting variables under a quasi-periodic representation. This expression is obtained numerically by frequency analysis. This allows us to characterise the equilibria of the system, and to distinguish the causes of their time variations.

 \par We show that, even without orbital eccentricity, the system can have complex behaviors, in particular when the core is highly 
flattened. In such a case, the polar motion is forced by several degrees and longitudinal librations appear. This is due to splitting of the equilibrium position of the polar motion. We also get a shift of the obliquity when the polar flattening of the core is small.

\keywords{Natural satellites \and Rotation \and Periodic Orbits \and Hamiltonian Systems \and Numerical Methods}
\end{abstract}

\section{Introduction}

\par Space missions like Galileo for the Jovian system or Cassini for the Saturnian one give us information on the internal structure
of the natural satellites, through their gravity fields (Anderson 2001 \cite{Anderson:2001}), observations of their surfaces 
(Porco et al. 2006 \cite{Porco:2006}) or measurements of their rotational states (Tiscareno et al. 2009 \cite{Tiscareno:2009},
Lorenz et al. 2008 \cite{Lorenz:2008}). It is known that the internal structure of a body influences its rotational dynamics, 
especially when this body is locked in a spin-orbit resonance, like the 1:1 resonance for most of the natural satellites of the 
Solar system, and the 3:2 resonance for Mercury. 

\par There are at least two ways to approach the modelisation of the interactions between the internal structure and the rotational 
dynamics. One way is to complexify the internal structure, taking account for instance of an atmosphere, a deformable crust, a 
subsurface ocean, an iron core\ldots in a simplified dynamical model that allows to consider only one degree of freedom 
(see e.g. Rambaux et al. (2011) \cite{Rambaux:2011a} for the longitudinal libration of satellites having an internal ocean, or 
Tokano et al. (2011) \cite{Tokano:2011} for the forcing of the polar motion of Titan due to its atmosphere). Another possibility is 
to consider a simple internal structure model (i.e. to assume the body to be rigid), in a full dynamical model considering several 
degrees of freedom (longitudinal motion, obliquity, and polar motion) like in (Henrard 2005 \cite{Henrard:2005,Henrard:2005a}). 

\par An evolution of this approach is to consider a two-layer body composed of a rigid mantle and an ellipsoidal fluid core in which
the flow is laminar and core-mantle interactions result in pressure coupling at the core-mantle boundary. This has been originally
written by Hough (1895) \cite{Hough:1895} and Poincar\'e (1910) \cite{Poincare:1910} (that is the reason why this model is sometimes 
called the Poincar\'e-Hough model), put in Hamiltonian form by Getino \& Ferr\'andiz (see e.g. \cite{Getino:1995a,Getino:1997}) under general 
assumptions, and by Touma \& Wisdom (2001) \cite{Touma:2001}, and recently used for Io (Henrard 2008 \cite{Henrard:2008}), Mercury 
(Noyelles et al. 2010 \cite{Noyelles:2010}) and the Moon (Meyer \& Wisdom 2011 \cite{Meyer:2011}). Another model exists, taking account of the 
elasticity of the mantle (Getino \& Ferr\'andiz 1995 \cite{Getino:1995}). This case will not be considered here.

\par In the case of the 1:1 spin-orbit resonance, the existing studies do not consider a wide range of internal structure parameter.
This paper aims at contributing to fill the gap to understand the behavior of the system for any size and shape (provided it is 
triaxial) of the core. The plan of the study is the following: after a description of the model, we present a systematic numerical 
study of the system with different sizes and shapes of the core, the considered body being an Io-like body on a low eccentricity orbit with a 
uniform nodal regression and constant inclination. Then, ``unusual'' behaviors are highlighted with analytical explanations.

\section{The model}

\par In the study of Henrard \cite{Henrard:2008}, the size of the core was not constrained, but its shape was assumed to be proportional to 
the whole Io. We here generalize this approach, in letting the shape parameters vary.

\subsection{Physical model}

\begin{figure}
\centering
\begin{tabular}{c}
\includegraphics[height=5cm,width=6.5cm]{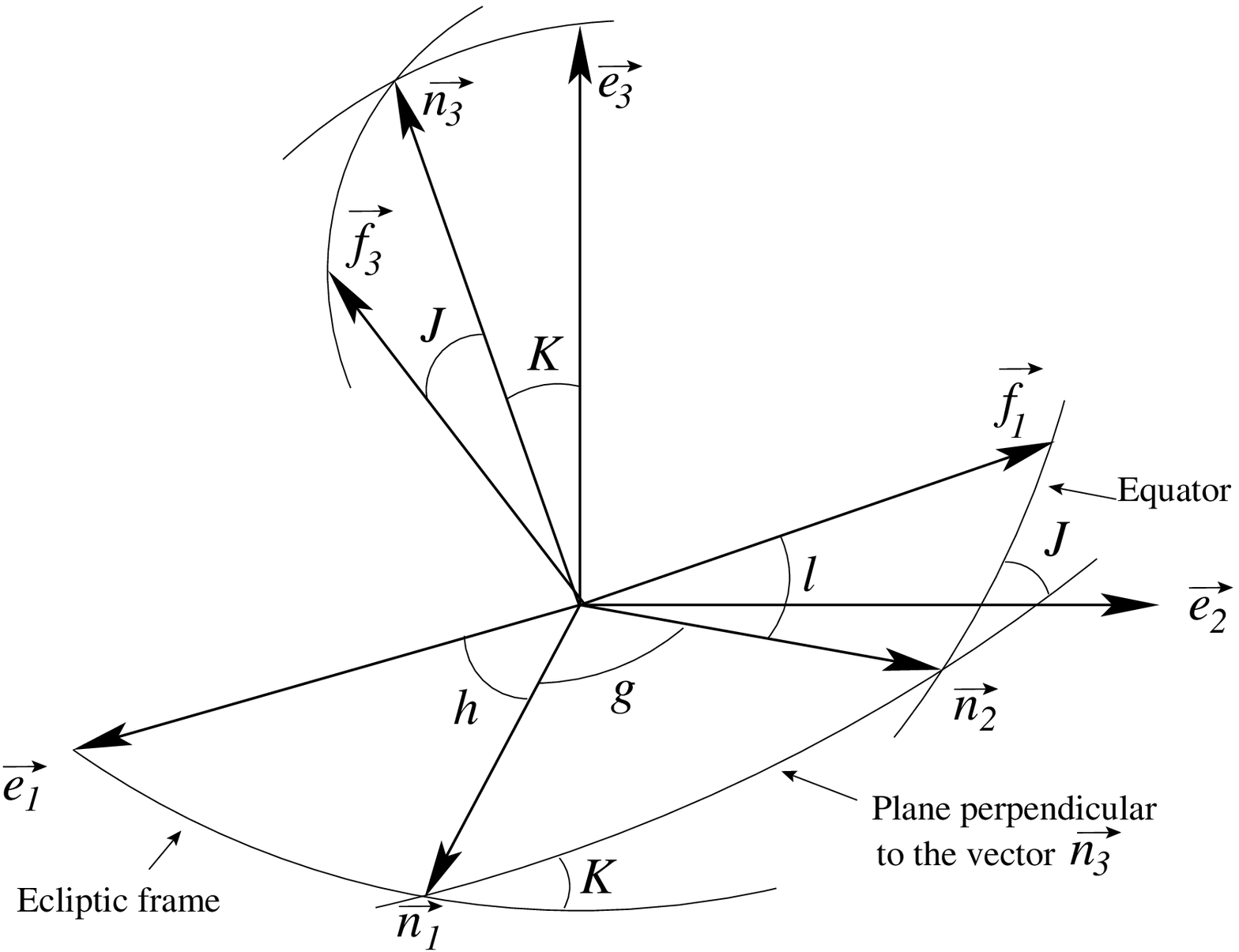} \\
\includegraphics[height=5cm,width=6.5cm]{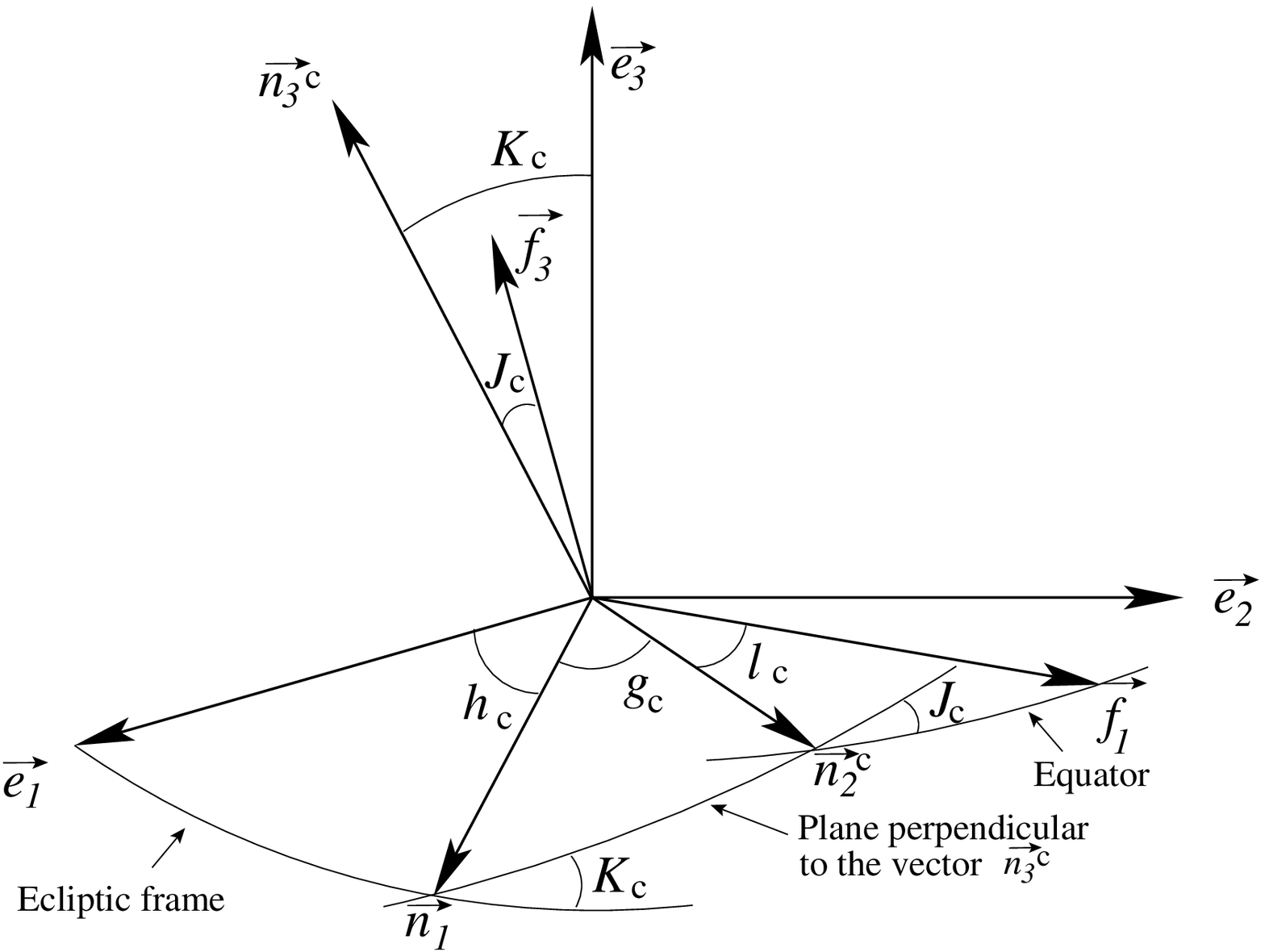}
\end{tabular}
\caption{In the upper panel we have 3 reference frames: one linked to the ecliptic plane ($\vec{e_1},\vec{e_2},\vec{e_3}$), 
another linked to the angular momentum $\vec N$ ($\vec{n_1},\vec{n_2},\vec{n_3}$), and the last one linked to 
the axes of inertia ($\vec{f_1},\vec{f_2},\vec{f_3}$) of the satellite. In the lower panel we have a similar configuration 
but instead of the angular momentum of the satellite, we have a reference frame linked to the angular momentum of a 
pseudo-core (defined later). We have the Euler angles $(h,K,g)$ positioning the vector $\vec{n_2}$ on the plane perpendicular 
to the angular momentum of the satellite and the Euler angles $(h_c,K_c,g_c)$ positioning the vector $\vec{n_2^c}$ on the 
plane perpendicular to the angular momentum of the pseudo-core. The angles $(l,J)$ and $(l_c,J_c)$ position the axis of 
least inertia. Note that $J_c$ is defined on the other side than $J$. \label{fig:rot}}
\end{figure}

\par Four references frames are considered (see Fig.\ref{fig:rot} \& \ref{fig:bigfig}). The first one, 
$(\vec{e_1},\vec{e_2},\vec{e_3})$ is assumed to be inertial for the rotational dynamics, it is in fact centered on 
the satellite and in translation with the inertial reference frame in which the motion of the satellite is defined. 
The second one, $(\vec{n^c_1},\vec{n^c_2},\vec{n^c_3})$ is linked to the angular momentum of a \emph{pseudo-core} that we 
define later, while the third one, i.e. $(\vec{n_1},\vec{n_2},\vec{n_3})$, is linked to the total angular momentum of 
the satellite. Finally, the last one, written as $(\vec{f_1},\vec{f_2},\vec{f_3})$, is rigidly linked to the principal axes 
of inertia of the satellite. In this last reference frame, the matrix of inertia of the satellite reads:

\begin{equation}
I=\left(\begin{array}{ccc}
A & 0 & 0 \\
0 & B & 0 \\
0 & 0 & C
\end{array}\right)
\label{equ:inertim}
\end{equation}
with $0<A\leq B \leq C$, while that of the core is:

\begin{equation}
I_c=\left(\begin{array}{ccc}
A_c & 0 & 0 \\
0 & B_c & 0 \\
0 & 0 & C_c
\end{array}\right),
\label{equ:inertic}
\end{equation}
in the same reference frame. So, the orientations of the mantle and the cavity are the same, a misalignment of their 
principal axes would require to consider the mantle as elastic, this is beyond the scope of the paper. This would in fact require
additional parameters related to the elasticity of the mantle, see e.g. (Getino \& Ferr\'andiz 1995 \cite{Getino:1995}).

\par As for the whole satellite, we have $0<A_c\leq B_c \leq C_c$. In this way, the principal moments of inertia of the mantle are 
respectively $A_m=A-A_c$, $B_m=B-B_c$ and $C_m=C-C_c$. The principal elliptical radii of the cavity are written respectively 
$a$, $b$, $c$, yielding

\begin{center}
$\begin{array}{ccccc}
A_c & = & \displaystyle\iiint(x_2^2+x_3^2)\rho\,dx_1\,dx_2\,dx_3 & = & \frac{M_c}{5}(b^2+c^2), \label{equ:Ac} \\
B_c & = & \displaystyle\iiint(x_1^2+x_3^2)\rho\,dx_1\,dx_2\,dx_3 & = &  \frac{M_c}{5}(a^2+c^2), \label{equ:Bc} \\
C_c & = & \displaystyle\iiint(x_1^2+x_2^2)\rho\,dx_1\,dx_2\,dx_3 & = &  \frac{M_c}{5}(a^2+b^2), \label{equ:Cc}
\end{array}$
\end{center}
where $\rho$ and $M_c$ are respectively the mass density and the mass of the fluid core, the quadrature being performed over the volume of the core.

\subsection{The kinetic energy of the system}

\par A Hamiltonian formulation of such a problem is usually composed of a kinetic energy and a disturbing potential, here the 
perturbation of the planet. Therefore, we consider every internal process, as the core-mantle interactions in our case, as part 
of the kinetic energy of the satellite. This section is widely inspired from (Henrard 2008 \cite{Henrard:2008}).

\par The components $(v_1,v_2,v_3)$ of the velocity field at the location $(x_1,x_2,x_3)$ inside the liquid core, in the frame of the
principal axes of inertia of the mantle, are assumed to be (Poincar\'e 1910 \cite{Poincare:1910}):

\begin{eqnarray}
v_1 & = & \left(\omega_2+\frac{a}{c}\nu_2\Big)x_3-\Big(\omega_3+\frac{a}{b}\nu_3\right)x_2, \label{equ:v1} \\
v_2 & = & \left(\omega_3+\frac{b}{a}\nu_3\Big)x_1-\Big(\omega_1+\frac{b}{c}\nu_1\right)x_3, \label{equ:v2} \\
v_3 & = & \left(\omega_1+\frac{c}{b}\nu_1\Big)x_2-\Big(\omega_2+\frac{c}{a}\nu_2\right)x_1, \label{equ:v3}
\end{eqnarray}
where $(\omega_1,\omega_2,\omega_3)$ are the components of the angular velocity of the mantle with respect to an inertial 
frame, and the vector of coordinates $(\nu_1,\nu_2,\nu_3)$ specifies the velocity field of the core with respect to the moving mantle. 
This vector is the velocity of a given fluid particle.
Here we assume that this velocity field $(\nu_1,\nu_2,\nu_3)$ depends only on the time $t$, and not on the spatial coordinates 
$(x_1,x_2,x_3)$. It implies that we have

\begin{equation}
  \vec{\nabla}\cdot\vec{v}=\frac{\partial v}{\partial x_1}+\frac{\partial v}{\partial x_2}+\frac{\partial v}{\partial x_3}=0,
\end{equation}
this equation is known as the continuity equation.

\par  The angular momentum of the core $\vec{N'_c}$ is obtained by:

\begin{equation}
\vec{N'_c}=\iiint_{core}(\vec{x}\times\vec{v})\rho\,dx_1\,dx_2\,dx_3
\label{equ:intNpc}
\end{equation}
and the result is:

\begin{equation}
\begin{split}
\vec{N'_c}= \frac{M_c}{5}\Bigg[\bigg(\frac{c}{b}\nu_1+\omega_1\bigg)b^2+\bigg(\frac{b}{c}\nu_1+\omega_1\bigg)c^2\Bigg]\vec{f_1} \\
 +\frac{M_c}{5}\Bigg[\bigg(\frac{c}{a}\nu_2+\omega_2\bigg)a^2+\bigg(\frac{a}{c}\nu_2+\omega_2\bigg)c^2\Bigg]\vec{f_2} \\
 +\frac{M_c}{5}\Bigg[\bigg(\frac{b}{a}\nu_3+\omega_3\bigg)a^2+\bigg(\frac{a}{b}\nu_3+\omega_3\bigg)b^2\Bigg]\vec{f_3}. \label{equ:Npc}
\end{split}
\end{equation}
We now set the following quantities:

\begin{center}
$\begin{array}{ccccc}
D_1 &=& \frac{2M_c}{5}bc &=& \sqrt{\big(A_c-B_c+C_c\big)\big(A_c+B_c-C_c\big)}, \\
D_2 &=& \frac{2M_c}{5}ac &=& \sqrt{\big(-A_c+B_c+C_c\big)\big(A_c+B_c-C_c\big)}, \\
D_3 &=& \frac{2M_c}{5}ab &=& \sqrt{\big(-A_c+B_c+C_c\big)\big(A_c-B_c+C_c\big)},
\end{array}$
\end{center}
that have the dimension of moments of inertia and can be seen as parameters of the core as $A_c$, $B_c$ and $C_c$, 
and we can write:

\begin{equation}
\vec{N'_c}=\big[A_c\omega_1+D_1\nu_1\big]\vec{f_1}+\big[B_c\omega_2+D_2\nu_2\big]\vec{f_2}+\big[C_c\omega_3+D_3\nu_3\big]\vec{f_3},
\label{equ:NPc2}
\end{equation}
while the angular momentum of the mantle is

\begin{equation}
\vec{N_m}=A_m\omega_1\vec{f_1}+B_m\omega_2\vec{f_2}+C_m\omega_3\vec{f_3},
\label{equ:Nman}
\end{equation}
and the total angular momentum of the satellite is

\begin{equation}
\vec{N}=\big[A\omega_1+D_1\nu_1\big]\vec{f_1}+\big[B\omega_2+D_2\nu_2\big]\vec{f_2}+\big[C\omega_3+D_3\nu_3\big]\vec{f_3}.
\label{equ:Ntot}
\end{equation}

\par The kinetic energy of the core is

\begin{equation}
T_c=\frac{1}{2}\iiint_{core}\rho v^2\,dx_1\,dx_2\,dx_3
\label{equ:tc}
\end{equation}
i.e.\footnote{we here correct a misprint present in Eq.13 of (Noyelles et al. 2010 \cite{Noyelles:2010})}

\begin{equation}
T_c=\frac{1}{2}\Big(A_c(\omega_1^2+\nu_1^2)+B_c(\omega_2^2+\nu_2^2)+C_c(\omega_3^2+\nu_3^2)+2D_1\omega_1\nu_1+2D_2\omega_2\nu_2+
2D_3\omega_3\nu_3\Big),
\label{equ:tc2}
\end{equation}
while the kinetic energy of the mantle $T_m$ is

\begin{equation}
T_m=\frac{1}{2}\vec{N_m}\cdot\vec{\omega}=\frac{A_m\omega_1^2+B_m\omega_2^2+C_m\omega_3^2}{2}.
\label{equ:Tm}
\end{equation}
From $T=T_m+T_c$ we finally deduce the kinetic energy of the satellite:

\begin{equation}
T=\frac{1}{2}\big(A\omega_1^2+B\omega_2^2+C\omega_3^2+A_c\nu_1^2+B_c\nu_2^2+C_c\nu_3^2+2D_1\omega_1\nu_1+2D_2\omega_2\nu_2+2D_3\omega_3\nu_3\big).
\label{equ:T}
\end{equation}

\par We can easily check the expressions of the partial derivatives, for instance

\begin{equation}
\frac{\partial T}{\partial \omega_1}  =  A\omega_1+D_1\nu_1  =  N_1
\label{equ:N1}
\end{equation}
or 

\begin{equation}
\frac{\partial T}{\partial \nu_1}  =  D_1\omega_1+A_c\nu_1  =  N_1^c,
\label{equ:N1c}
\end{equation}
where $N_i$ are the components of the total angular momentum. $N_i^c$ are not the components of the angular momentum of the core 
but are close to it for a cavity close to spherical. We have, for instance for the first component:

\begin{equation}
N_1^c-N_1'^c=(A_c-D_1)(\omega_1-\nu_1)=\frac{M_c}{5}(c-b)^2(\omega_1-\nu_1),
\label{equ:depart}
\end{equation}
so the difference is of the second order in departure from the sphericity. From now on, we call \textit{angular momentum of 
the pseudo-core} the vector $\vec{N^c}=N_1^c\vec{f_1}+N_2^c\vec{f_2}+N_3^c\vec{f_3}$.

\par With these notations, the Poincar\'e-Hough's equations of motion, for the system mantle-core in the absence of external 
torque, are (see e.g. Eq.15 in \cite{Touma:2001} or \cite{Henrard:2008}):

\begin{eqnarray}
\frac{d\vec{N}}{dt} & = & \vec{N} \times \vec{\nabla}_{\vec{N}}\mathcal{T}, \label{equ:ph1} \\
\frac{d\vec{N^c}}{dt} & = & \vec{N^c} \times \vec{\nabla}_{-\vec{N^c}}\mathcal{T}, \label{equ:ph2}
\end{eqnarray}
with
\begin{equation}
\label{equ:gradient}
\vec{\nabla}_{\vec{N}}\mathcal{T}=\frac{\partial \mathcal{T}}{\partial N_1}\vec{f_1}+\frac{\partial \mathcal{T}}{\partial N_2}\vec{f_2}+\frac{\partial \mathcal{T}}{\partial N_3}\vec{f_3},
\end{equation}
and
\begin{equation}
\label{equ:gradient2}
\vec{\nabla}_{-\vec{N^c}}\mathcal{T}=-\frac{\partial \mathcal{T}}{\partial N_1^c}\vec{f_1}-\frac{\partial \mathcal{T}}{\partial N_2^c}\vec{f_2}-\frac{\partial \mathcal{T}}{\partial N_3^c}\vec{f_3}.
\end{equation}
Here $\mathcal{T}$ is the kinetic energy expressed in terms of the components of the vectors $\vec{N}$ and $\vec{N^c}$, i.e.

\begin{equation}
\begin{split}
\mathcal{T}=\frac{1}{2\alpha}\big(A_cN_1^2+A(N_1^c)^2-2D_1N_1N_1^c\big)+\frac{1}{2\beta}\big(B_cN_2^2+B(N_2^c)^2-2D_2N_2N_2^c\big) \\
+\frac{1}{2\gamma}\big(C_cN_3^2+C(N_3^c)^2-2D_3N_3N_3^c\big)
\end{split}
\end{equation}
with $\alpha=AA_c-D_1^2$, $\beta=BB_c-D_2^2$ and $\gamma=CC_c-D_3^2$.

\subsection{The Hamiltonian}

\subsubsection{The rotational kinetic energy}

\par We assume that the cavity and the satellite are almost spherical, this allows us to introduce the four small parameters 
$\epsilon_i$:

\begin{eqnarray}
\epsilon_1 &=& \frac{2C-A-B}{2C}=J_2\frac{MR^2}{C}, \label{equ:eps1} \\
\epsilon_2 &=& \frac{B-A}{2C}=2C_{22}\frac{MR^2}{C}, \label{equ:eps2} \\
\epsilon_3 &=& \frac{2C_c-A_c-B_c}{2C_c}, \label{equ:eps3} \\
\epsilon_4 &=& \frac{B_c-A_c}{2C_c}, \label{equ:eps4}
\end{eqnarray}
where $M$ is the mass of our body and $R$ its mean radius, and also the parameter $\delta=C_c/C$, i.e. the ratio between the polar inertial momentum of the core and of the satellite. $\epsilon_1$ represents the polar flattening of the satellite, while $\epsilon_2$ is its equatorial ellipticity. $\epsilon_3$ and $\epsilon_4$ have the same meaning for the cavity. If we assume the core of the satellite to be spherical, we should take $\epsilon_3=\epsilon_4=0$, while $\epsilon_4=0$ represents an axisymmetric cavity. Henrard \cite{Henrard:2008} considered that the ellipsoid of inertia of the core and the mantle were proportional, the mathematical formulation was $\epsilon_3=\epsilon_1$ and $\epsilon_4=\epsilon_2$.

\par
\begin{figure}
\centering
\includegraphics[width=7cm]{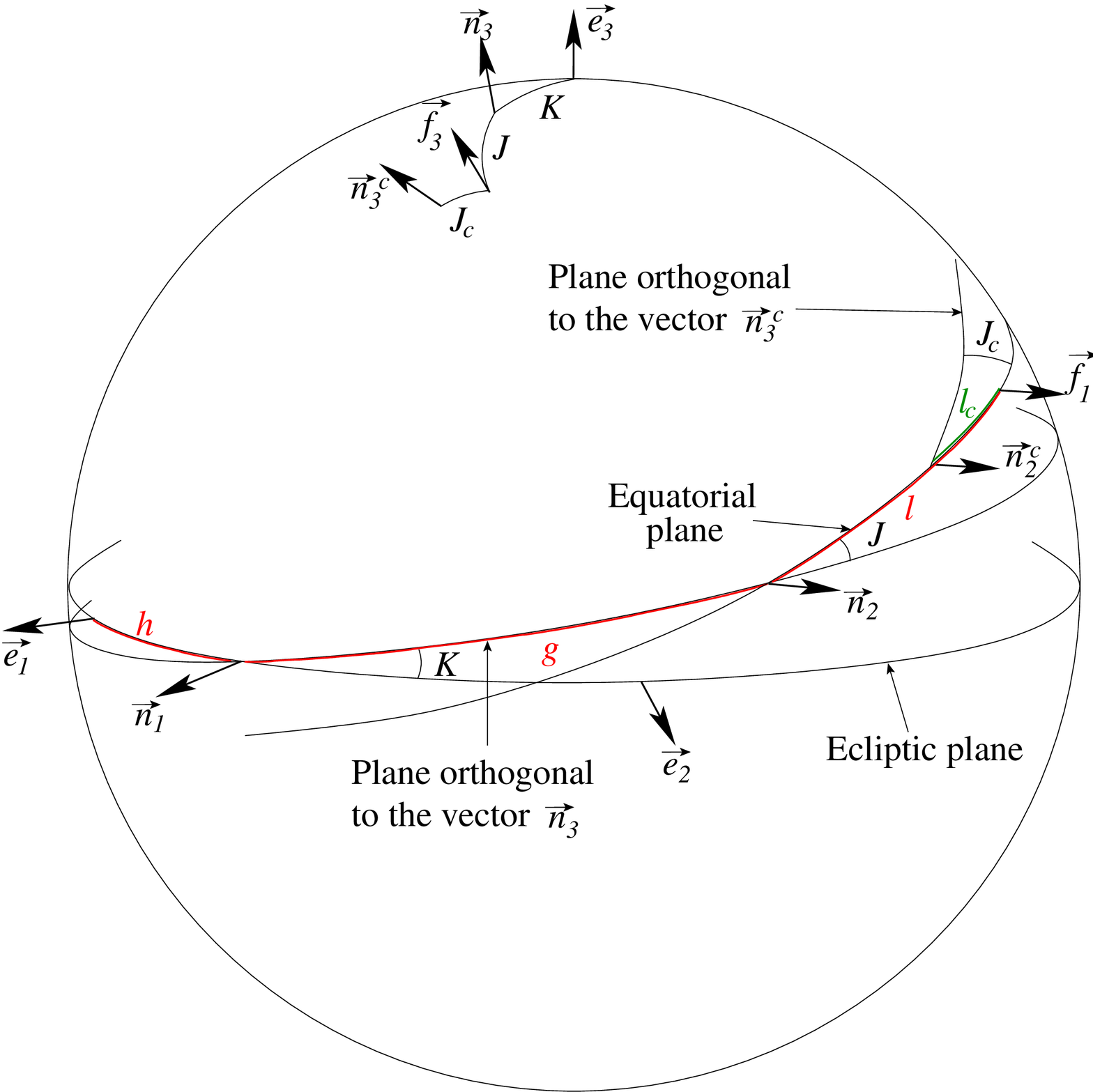}
\caption{The four reference frames gathered in the same view. The angles $(h,K)$ position the plane orthogonal to the angular 
momentum $\vec N$. The Euler angles $(g,J,l)$ locate the axis of least inertia and the body frame 
$(\vec{f_1},\vec{f_2},\vec{f_3})$. The angles $(J_c,l_c)$ place the angular momentum of the pseudo-core with respect to the 
axis of least inertia $f_1$.\label{fig:bigfig}}
\end{figure}

 We now introduce the two sets of Andoyer's variables \cite{Andoyer:1926}, $(l,g,h,L,G,H)$ and $(l_c,g_c,h_c,L_c,G_c,H_c)$, 
related respectively to the whole satellite and to its core. The angles $(h,K,g)$ are the Euler angles of the vector $\vec{n_2}$, 
node of the equatorial plane over the plane perpendicular to the angular momentum $\vec{N}$, the angles $(J,l)$ position the axis 
of least inertia $\vec{f_1}$ with respect to $\vec{n_2}$. Correspondingly the angles $(h_c,K_c,g_c)$ are the Euler angles of the 
vector $\vec{n^c_2}$, node of the equatorial plane over the plane perpendicular to the angular momentum of the 
pseudo-core $\vec{N_c}$, and $(J_c,l_c)$ position the axis of least inertia with respect to $\vec{n^c_2}$. 
The Figure \ref{fig:bigfig} shows a schematic view of all the reference frames and relevant angles. The variables 
are $(h,g,l)$ and $(h_c,g_c,l_c)$ and the corresponding momenta ($H=N\cos K$, $G=N$, $L=N\cos J$) and 
($H_c=N^c \cos K_c$, $G_c=N^c$, $L_c=N^c \cos J_c$). Expressed in Andoyer's variables the components of 
$\vec{N}$ and $\vec{N^c}$ are:

\begin{center}
$\begin{array}{lll}
N_1=\sqrt{G^2-L^2}\sin l, & \hspace{2cm} & N_1^c=\sqrt{G_c^2-L_c^2}\sin l_c, \\
N_2=\sqrt{G^2-L^2}\cos l, & \hspace{2cm} & N_2^c=\sqrt{G_c^2-L_c^2}\cos l_c, \\
N_3=L, & \hspace{2cm} & N_3^c=L_c. \\
\end{array}$
\end{center}

\par We can now straightforwardly derive the Hamiltonian $\mathcal{H}_1$ of the free rotation of the satellite, using 
Andoyer's variables and changing the sign of $\vec{N^c}$ to take the minus sign of the Poincar\'e-Hough equations into 
account (Eq.\ref{equ:ph2}). We also linearize the Hamiltonian with respect to the small parameters $\epsilon_i$ (their orders 
of magnitude being about $10^{-5}$, see Tab.\ref{tab:pseudoIo}), and get:

\begin{eqnarray}
\mathcal{H}_0&=&\frac{1}{2C(1-\delta)}\bigg(G^2+\frac{G_c^2}{\delta}+2\sqrt{(G^2-L^2)(G_c^2-L_c^2)}\cos(l-l_c)+2LL_c\bigg) \nonumber \\
&+&\frac{\epsilon_1}{2C(1-\delta)^2}\bigg(G^2-L^2+G_c^2-L_c^2+2\sqrt{(G^2-L^2)(G_c^2-L_c^2)}\cos(l-l_c)\bigg) \nonumber \\
&-&\frac{\epsilon_2}{2C(1-\delta)^2}\bigg((G^2-L^2)\cos(2l)+(G_c^2-L_c^2)\cos(2l_c) \nonumber \\
 & & +2\sqrt{(G^2-L^2)(G_c^2-L_c^2)}\cos(l+l_c)\bigg) \nonumber \\
&-&\frac{\epsilon_3}{2C(1-\delta)^2}\bigg(\delta (G^2-L^2)+(G_c^2-L_c^2)(2-\frac{1}{\delta}) \nonumber \\
 & & +2\delta\sqrt{(G^2-L^2)(G_c^2-L_c^2)}\cos(l-l_c)\bigg) \nonumber \\
&+&\frac{\epsilon_4}{2C(1-\delta)^2}\bigg(\delta(G^2-L^2)\cos(2l)+(G_c^2-L_c^2)(2-\frac{1}{\delta})\cos(2l_c) \nonumber \\
& & +2\delta\sqrt{(G^2-L^2)(G_c^2-L_c^2)}\cos(l+l_c)\bigg). \label{equ:HG}
\end{eqnarray}

\par We now introduce the following canonical change of variables, of multiplier $\frac{1}{nC}$, $n$ being the mean orbital 
motion of the satellite:

\begin{equation}\label{eq:chvar}
\begin{array}{lll}
p=l+g+h, & \hspace{2cm} & P=\frac{G}{nC}, \\
r=-h, & \hspace{2cm} & R=P(1-\cos K), \\
\xi_1=-\sqrt{2P(1-\cos J)}\sin l, & \hspace{2cm} & \eta_1=\sqrt{2P(1-\cos J)}\cos l, \\
p_c=-l_c+g_c+h_c, & \hspace{2cm} & P_c=\frac{G_c}{nC}, \\
r_c=-h_c, & \hspace{2cm} & R_c=P_c(1-\cos K_c), \\
\xi_2=\sqrt{2P_c(1+\cos J_c)}\sin l_c, & \hspace{2cm} & \eta_2=\sqrt{2P_c(1+\cos J_c)}\cos l_c. \\
\end{array} \\
\end{equation}

\par The first three lines of this new set of variables and associated moments are related to the whole body, while the last three ones are related to the pseudo-core. $P$ is the normalized norm of the angular momentum, it should be close to $1$ at the spin-orbit resonance. Since the obliquity $K$ is small, we have $R\propto K^2$, i.e. this is a small quantity related to the obliquity of the body. The quantities $(\xi_1,\eta_1)$ are related to the polar motion of the body, i.e. the angle $J$ between the geometrical polar axis and the angular momentum, while $l$ is the precession angle associated. We can note that $\xi_1$ and $\eta_1$ are always defined, while $l$ is not defined when $J=0$. The last three lines have basically the same meaning for the pseudo-core. We will see later that the degree of freedom $(r_c,R_c)$ is in fact not involved in the dynamics of this model, and that $p_c$ is not involved either, letting the norm of the angular momentum of the pseudo-core $P_c$ to be a constant. So, we can consider that the rotational dynamics of our body has 4, and not 6, degrees of freedom. 

\par In order to be consistent with the minus sign in the equations and before $l_c$, the amplitude of the wobble of the 
pseudo-core $J_c$ has to be replaced by $\pi-J_c$. In this way, we have $L_c=G_c\cos(\pi-J_c)=-G_c\cos(J_c)$. In this new set of 
variables, we have

\begin{center}
$\begin{array}{lll}
N_1=-nC\sqrt{P^2-\Big(P-\frac{\xi_1^2+\eta_1^2}{2}\Big)^2}\frac{\xi_1}{\xi_1^2+\eta_1^2}, & \hspace{0.6cm} & N_1^c=nC\sqrt{P_c^2-\Big(\frac{\xi_2^2+\eta_2^2}{2}-P_c\Big)^2}\frac{\xi_2}{\xi_2^2+\eta_2^2}, \\
N_2=nC\sqrt{P^2-\Big(P-\frac{\xi_1^2+\eta_1^2}{2}\Big)^2}\frac{\eta_1}{\xi_1^2+\eta_1^2}, & \hspace{0.6cm} & N_2^c=nC\sqrt{P_c^2-\Big(\frac{\xi_2^2+\eta_2^2}{2}-P_c\Big)^2}\frac{\eta_2}{\xi_2^2+\eta_2^2}, \\
N_3=nC\Big(P-\frac{\xi_1^2+\eta_1^2}{2}\Big), & \hspace{0.6cm} & N_3^c=nC\Big(\frac{\xi_2^2+\eta_2^2}{2}-Pc\Big), \\
\end{array}$ \\
\end{center}
and the Hamiltonian of the free rotational motion becomes, after division by $nC$: 
\begin{eqnarray}
\mathcal{H}_1&=&\frac{n}{2(1-\delta)}\Bigg(P^2+\frac{P_c^2}{\delta}+2\sqrt{\Big(P-\frac{\xi_1^2+\eta_1^2}{4}\Big)\Big(P_c-\frac{\xi_2^2+\eta_2^2}{4}\Big)}\big(\eta_1\eta_2-\xi_1\xi_2\big) \nonumber \\
 & & +2\Big(P-\frac{\xi_1^2+\eta_1^2}{2}\Big)\Big(\frac{\xi_2^2+\eta_2^2}{2}-P_c\Big)\Bigg) \nonumber \\
%& & \nonumber \\
&+&\frac{n\epsilon_1}{2(1-\delta)^2}\Bigg(P_c^2-\Big(\frac{\xi_2^2+\eta_2^2}{2}-P_c\Big)^2+P^2-\Big(P-\frac{\xi_1^2+\eta_1^2}{2}\Big)^2 \nonumber \\
&+& 2\sqrt{\Big(P-\frac{\xi_1^2+\eta_1^2}{4}\Big)\Big(P_c-\frac{\xi_2^2+\eta_2^2}{4}\Big)}\big(\eta_1\eta_2-\xi_1\xi_2\big)\Bigg) \nonumber \\
%& & \nonumber \\
&+&\frac{n\epsilon_2}{2(1-\delta)^2}\Bigg(\frac{1}{4}\big(4P-\xi_1^2-\eta_1^2\big)\big(\xi_1^2-\eta_1^2\big)+\frac{1}{4}\big(4P_c-\xi_2^2-\eta_2^2\big)\big(\xi_2^2-\eta_2^2\big) \nonumber \\
&-&2\sqrt{\Big(P-\frac{\xi_1^2+\eta_1^2}{4}\Big)\Big(P_c-\frac{\xi_2^2+\eta_2^2}{4}\Big)}\big(\eta_1\eta_2+\xi_1\xi_2\big)\Bigg) \nonumber \\
%& & \nonumber \\
&-&\frac{n\epsilon_3}{2(1-\delta)^2}\Bigg(\delta \Big(P^2-\Big(P-\frac{\xi_1^2+\eta_1^2}{2}\Big)^2\Big)+\Big(P_c^2-(\frac{\xi_2^2+\eta_2^2}{2}-P_c\Big)^2\Big)\Big(2-\frac{1}{\delta}\Big) \nonumber \\
&+& 2\delta\sqrt{\Big(P-\frac{\xi_1^2+\eta_1^2}{4}\Big)\Big(P_c-\frac{\xi_2^2+\eta_2^2}{4}\Big)}\big(\eta_1\eta_2-\xi_1\xi_2\big)\Bigg) \nonumber \\
%& & \nonumber \\
&+&\frac{n\epsilon_4}{2(1-\delta)^2}\Bigg(\frac{\delta}{4}\big(4P-\xi_1^2-\eta_1^2\big)\big(\eta_1^2-\xi_1^2\big)+\Big(2-\frac{1}{\delta}\Big) \frac{1}{4}\big(4P_c-\xi_2^2-\eta_2^2\big)\big(\eta_2^2-\xi_2^2\big)  \nonumber \\
&+ & 2\delta\sqrt{\Big(P-\frac{\xi_1^2+\eta_1^2}{4}\Big)\Big(P_c-\frac{\xi_2^2+\eta_2^2}{4}\Big)}\big(\eta_1\eta_2+\xi_1\xi_2\big)\Bigg). \label{equ:HG3}
\end{eqnarray}

\par Finally, in order to get an easy-to-read formula, we can develop this Hamiltonian up to the second order in 
($\xi_1$, $\xi_2$, $\eta_1$, $\eta_2$) to get:

\begin{eqnarray}
\mathcal{H}_1&\approx&\frac{n}{2(1-\delta)}\Bigg(P^2+\frac{P_c^2}{\delta}+2\sqrt{PP_c}\big(\eta_1\eta_2-\xi_1\xi_2\big) \nonumber \\
 & & +2\Big(P\frac{\xi_2^2+\eta_2^2}{2}+P_c\frac{\xi_1^2+\eta_1^2}{2}-PP_c\Big)\Bigg) \nonumber \\
& & \nonumber \\
&+&\frac{n\epsilon_1}{2(1-\delta)^2}\Bigg(P\big(\xi_1^2+\eta_1^2\big)+P_c\big(\xi_2^2+\eta_2^2\big)+2\sqrt{PP_c}\big(\eta_1\eta_2-\xi_1\xi_2\big)\Bigg) \nonumber \\
& & \nonumber \\
&+&\frac{n\epsilon_2}{2(1-\delta)^2}\Bigg(P\big(\xi_1^2-\eta_1^2\big)+P_c\big(\xi_2^2-\eta_2^2\big)-2\sqrt{PP_c}\big(\eta_1\eta_2+\xi_1\xi_2\big)\Bigg) \label{equ:HG4} \\
& & \nonumber \\
&-&\frac{n\epsilon_3}{2(1-\delta)^2}\Bigg(\delta P\big(\xi_1^2+\eta_1^2\big)+\Big(2-\frac{1}{\delta}\Big)P_c \big(\xi_2^2+\eta_2^2\big)+2\delta\sqrt{PP_c}\big(\eta_1\eta_2-\xi_1\xi_2\big)\Bigg) \nonumber \\
& & \nonumber \\
&+&\frac{n\epsilon_4}{2(1-\delta)^2}\Bigg(\delta P\big(\eta_1^2-\xi_1^2\big)+\Big(2-\frac{1}{\delta}\Big)P_c\big(\eta_2^2-\xi_2^2\big)+2\delta\sqrt{PP_c}\big(\eta_1\eta_2+\xi_1\xi_2\big)\Bigg). \nonumber
\end{eqnarray}
This is in fact a third-order development since the powers in $(\xi_1,\xi_2,\eta_1,\eta_2)$ are even. In the forthcoming computations, this last approximation has not been used, the equations we have propagated deriving from the Hamiltonian (\ref{equ:HG3}).

\subsubsection{The gravitational potential}

To compute the gravitational potential due to the parent planet on its satellite, we must first obtain the coordinates $x$, $y$, 
and $z$ of the unit vector pointing to the planet in the reference frame linked to the principal axes of inertia $(\vec{f_1},\vec{f_2},\vec{f_3})$, from its coordinates in the inertial frame $x_i$, $y_i$ and $z_i$. Five rotations are to be performed:
\begin{equation}
\left(\begin{array}{c}
x \\
y \\
z
\end{array}\right)
=R_3(-l)R_1(-J)R_3(-g)R_1(-K)R_3(-h)\left(\begin{array}{c}
x_i \\
y_i \\
z_i
\end{array}\right)
\label{equ:passage}
\end{equation}
with $x_i$, $y_i$, $z_i$ depending on the mean longitude $\lambda_o$, the longitude of the ascending node $\ascnode_o$, the longitude of 
the perihelion $\varpi_o$, the inclination $i$, and the eccentricity $e$.\\
The rotation matrices are defined by

\begin{equation}
R_3(\phi)=\left(\begin{array}{ccc}
\cos\phi & -\sin\phi & 0 \\
\sin\phi & \cos\phi & 0 \\
0 & 0 & 1
\end{array}\right),\qquad
R_1(\phi)=\left(\begin{array}{ccc}
1 & 0 & 0 \\
0 & \cos\phi & -\sin\phi \\
0 & \sin\phi & \cos\phi 
\end{array}\right).
\label{equ:r3}
\end{equation}
The gravitational potential then reads:
\begin{equation}
V_1(\lambda_o,l,g,h,J,K)=-\frac{3}{2}C\frac{\mathcal{G}M_p}{d^3}\big(\epsilon_1(x^2+y^2)+\epsilon_2(x^2-y^2)\big)
\label{equ:pull1}
\end{equation}
where $\mathcal{G}$ is the gravitational constant, $M_p$ the mass of the perturber, i.e. Jupiter for Io, $(x,y,z)$ the unit vector pointing at the 
perturber in the frame $(\vec{f_1},\vec{f_2},\vec{f_3})$, such that $x^2+y^2+z^2=1$, while $d$ is the distance planet-satellite. \\
Let us note that unlike Henrard \cite{Henrard:2008}, we consider that the perturbation is applied to the whole satellite and not only to 
its mantle, this issue is addressed in Noyelles et al. (2010 \cite{Noyelles:2010}). \\
From the variables $x$, $y$ and $z$, it is easy to introduce the set of variables defined in (Eq. \ref{eq:chvar}). We also modify 
the moment $\Lambda_o$ associated with $\lambda_o$ (that appears in the expressions of $x$ and $y$) in such way that all our variables 
are now canonical with multiplier $1/nC$ and our gravitational potential becomes (after division by $nC$)
\begin{equation}
\mathcal{H}_2(\lambda_o,p,P,r,R,\xi_1,\eta_1)=-\frac{3}{2}\frac{\mathcal{G}M_p}{nd^3}\big(\epsilon_1(x^2+y^2)+\epsilon_2(x^2-y^2)\big).
\label{equ:pull2}
\end{equation}

\par Finally, we use the formulae (\ref{equ:HG3}) and (\ref{equ:pull2}) to get the Hamiltonian of the system:

\begin{equation}
\mathcal{H}=\mathcal{H}_1(P,\xi_1,\eta_1,\xi_2,\eta_2)+\mathcal{H}_2(\lambda_o,p,P,r,R,\xi_1,\eta_1).
\label{equ:hamiltout}
\end{equation}
The four degrees of freedom of this Hamiltonian are the spin ($p$, $P$), the obliquity ($r$, $R$), the wobble of the whole 
body ($\xi_1$, $\eta_1$) and the wobble of the core ($\xi_2$, $\eta_2$).

\subsubsection{Evaluating $P_c$}

\par Since the variable $p_c$, spin angle of the pseudo-core, does not appear explicitly in the Hamiltonian of the system, its associated momentum $P_c$
, norm of the angular momentum of the pseudo-core is not ruled by the Hamilton equations. So, it can be either a constant, or a time varying input as is the 
orbital motion of the system. 
We here choose to set $P_c=\delta=C_c/C$, the mean value of $P$ being very close to $1$ as our pseudo-Io is in 1:1 spin-orbit 
resonance. So, we assume a kind of equipartition of the norm of the angular momentum between the core and the mantle. 

\par An exact equipartition would be $P_c(t)=\delta P(t)$, meaning that the fluid would follow every fluctuation of the orbital velocity of our 
pseudo-Io. It would mean that the fluid follows the longitudinal librations of the mantle, as if it were rigid. In such a case, the 
amplitude of the longitudinal librations would not be affected by the presence of an at least partially liquid core. Observations of 
such librations for Mercury (Margot et al. 2007 \cite{Margot:2007}) and the Moon (Koziel 1967 \cite{Koziel:1967}, Williams et al.
1973 \cite{Williams:1973}) support the assumption that the longitudinal librations are the response of the solid mantle (and not of 
the full body) to variations of the orbital velocity of the body. That is the reason why we consider a constant value for $P_c$, 
that results from a kind of rough averaging of $P$.

\par While this model describes the rigid dynamics of a body having a fluid core, we must not forget that real bodies on which this model could be applied have a 
viscous fluid core. We here discuss the relevance of our assumptions on $P_c$ for these bodies. From a physical point of view, the reason for the decoupling between
the fluid and the mantle is a low viscosity of the fluid. At the core-mantle boundary
(CMB), the no-slip condition imposes that the velocity field follows the mantle. So, there is a thin turbulent layer close to this 
boundary, known as the Ekman layer, in which the velocity field evolves continuously from the no-slip condition at the boundary to 
the one satisfying $P_c=\delta$. The typical thickness of the Ekman layer is $d=\sqrt{\nu/\Omega}$ (Greenspan 1968 
\cite{Greenspan:1968}), $\nu$ being the kinematic viscosity and $\Omega=n$ the spin frequency of the fluid. Usually a kinematic 
viscosity $\nu=10^{-6}m^2/s$ is considered at the core-mantle boundary because it is consistent with a Fe/Fe-S composition 
(see e.g. Kerswell 1998 \cite{Kerswell:1998}), what 
yields $d=0.16$ m. A viscosity of $36m^2/s$ is necessary for the thickness of the Ekman layer to reach 1 km. In fact, the viscosity 
is expected to increase with the depth under the CMB, since molten, and even rigid iron, should be concentrated at the inner core 
(see e.g. Rutter et al. \cite{Rutter:2002}). We anecdotally recall the extremum of viscosity of the pitch derived from the pitch drop 
experiment set up in 1927 at the University of Queensland, Australia (Edgeworth et al. 1984 \cite{Edgeworth:1984}), i.e. 
$\nu=(2.09\times10^5\pm4.6\times10^4)m^2/s$.

\subsection{Link with the Navier-Stokes equation}

\par As said in the introduction, there are at least two ways to approach the interactions between the internal structure and the rotational dynamics. One is to complexify the internal structure in considering only one degree of freedom, and the other one is to consider several dynamical degrees of freedom (4 in this study) with a quite simple internal structure. We must keep in mind that these 2 very different approaches aim at studying the same bodies. A complete study of the core dynamics would require to consider the Navier-Stokes equation, we here make a link with this equation to help in the interpretation of our model from a physical point of view.

\par The dynamics of a particle of fluid is often assumed to be ruled by the well-known Navier-Stokes equation, we give here its 
expression as given in (Greenspan 1968 \cite{Greenspan:1968}):

\begin{equation}
\label{eq:navierstokes}
  \frac{\partial}{\partial t}\vec{q}+(\vec{q}\cdot\vec{\nabla})\vec{q}+2\vec{\Omega}\times\vec{q}=
-\frac{1}{\rho}\vec{\nabla}p-\nu\vec{\nabla}\times(\vec{\nabla}\times\vec{q})-\vec{r}\times\frac{d\vec{\Omega}}{dt},
\end{equation}
with

\begin{itemize}

  \item $\vec{q}$: particle velocity measured in a rotating system

  \item $\vec{\Omega}$: angular velocity of the rotating system, its coordinates being $(\omega_1,\omega_2,\omega_3)$

  \item $\rho$: density of the fluid

  \item $\vec{r}$: position of the particle

  \item $p=P+\rho\mathcal{U}-\frac{\rho}{2}(\vec{\Omega}\times\vec{r})\cdot(\vec{\Omega}\times\vec{r})$: the reduced pressure, where 
$P$ is the pressure of the fluid, and $\mathcal{U}$ an exterior potential,

  \item $\nu$: kinematic viscosity of the fluid.

\end{itemize}
In our case we have

\begin{equation}
\vec{q}=\left(\begin{array}{c}
(a/c)\nu_2x_3-(a/b)\nu_3x_2 \\
(b/a)\nu_3x_1-(b/c)\nu_1x_3 \\
(c/b)\nu_1x_2-(c/a)\nu_2x_1
\end{array}\right).
\label{equ:vecq}
\end{equation}

In an over-simplified case where we neglect the viscosity $\nu$, the convective acceleration $(\vec{q}\cdot\vec{\nabla})\vec{q}$ and
the reduced pressure $p$, the formula (\ref{eq:navierstokes}) reads:

\begin{equation}
  \label{eq:navierstokes2}
  \frac{\partial}{\partial t}\vec{q}+2\vec{\Omega}\times\vec{q}=\vec{0},
\end{equation}
i.e.

\begin{eqnarray}
  \frac{d\nu_1}{dt}+2(\omega_2\nu_3-\omega_3\nu_2) & = & 0, \nonumber \\
  \frac{d\nu_2}{dt}+2(\omega_3\nu_1-\omega_1\nu_3) & = & 0, \label{eq:nv4} \\
  \frac{d\nu_3}{dt}+2(\omega_1\nu_2-\omega_2\nu_1) & = & 0. \nonumber
\end{eqnarray}
For comparison, the formula (\ref{equ:ph1}) reads:

\begin{eqnarray}
  A\frac{d\omega_1}{dt}+D_1\frac{d\nu_1}{dt} & = & (B\omega_2+D_2\nu_2)\omega_3-(C\omega_3+D_3\nu_3)\omega_2, \nonumber \\
  B\frac{d\omega_2}{dt}+D_2\frac{d\nu_2}{dt} & = & (C\omega_3+D_3\nu_3)\omega_1-(A\omega_1+D_1\nu_1)\omega_3, \label{eq:nv5} \\
  C\frac{d\omega_3}{dt}+D_3\frac{d\nu_3}{dt} & = & (A\omega_1+D_1\nu_1)\omega_2-(B\omega_2+D_2\nu_2)\omega_1. \nonumber
\end{eqnarray}
The systems of equations (\ref{eq:nv4}) and (\ref{eq:nv5}) present some similarities, the main difference being that the moments of 
inertia are involved in Eq.\ref{eq:nv5}. They should be in fact considered as global equations (i.e. considering the whole volume of
fluid), while the Eq.\ref{eq:nv4} is a local form, considering an individual fluid particle.

\par The reader can find another formulation of these equations in (Rambaux et al. 2007 \cite{Rambaux:2007}).

\section{A numerical study}

\subsection{The algorithm}

\par As shown in Henrard \cite{Henrard:2008}, the proper frequency associated with the core, i.e. the free core nutation, is close to 
the spin period of the considered body. For a synchronous satellite, this period is also the orbital period, so we have a proximity
between a proper frequency of the problem and a forcing period. As a consequence, a perturbative approach will meet difficulties to 
converge because of small divisors. Such a problem has already been encountered in (Noyelles et al. 2010 \cite{Noyelles:2010}). That is 
the reason why we prefer a full numerical study, consisting of a numerical integration of the equations derived from the Hamiltonian 
(\ref{equ:hamiltout}), and a frequency analysis of the solutions of the problem. The frequency analysis algorithm we use is widely
inspired from NAFF (see Laskar 1993 \cite{Laskar:1993} for the method, and Laskar 2005 \cite{Laskar:2005} for the convergence proofs), 
with a refinement suggested by Champenois (1998 \cite{Champenois:1998}) consisting in iterating the process to enhance the accuracy 
of the determination.

\par The basic idea of the frequency analysis is to consider that a complex variable of the problem $x(t)$ is quasi-periodic, i.e. 
can be expressed as a, a priori infinite, sum of a converging trigonometric series like

\begin{equation}
 x(t)=\sum_{n=0}^{\infty} A_n \exp\left(\imath \nu_nt\right)
\end{equation}
where $A_n$ are constant complex amplitudes, and $\nu_n$ constant frequencies, with 

\begin{equation}
 x(t)\approx\sum_{n=0}^{N} A_n^{\bullet} \exp\left(\imath \nu_n^{\bullet}t\right),
\label{eq:naffc}
\end{equation}
the bullet meaning that the coefficients have been numerically determined. A detailed description of the algorithm is given in 
appendix. In the case of a real variable, the Eq.\ref{eq:naffc} becomes

\begin{equation}
 x(t)\approx\sum_{n=0}^{N} A_n^{\bullet} \cos\left(\nu_n^{\bullet}t+\phi_n^{\bullet}\right),
\label{eq:naffrc}
\end{equation}
or

\begin{equation}
 x(t)\approx\sum_{n=0}^{N} A_n^{\bullet} \sin\left(\nu_n^{\bullet}t+\phi_n^{\bullet}\right),
\label{eq:naffrs}
\end{equation}
where the amplitudes are now real, and the $\phi_n^{\bullet}$ are real phases expressed with the counterclockwise convention, previously included in the complex amplitudes.

\par The rotation of a synchronous satellite is reputed to have reached an equilibrium state, known as Cassini State 1 (see e.g. 
Cassini 1693 \cite{Cassini:1693}, Peale 1969 \cite{Peale:1969}, and Bouquillon et al. 2003 \cite{Bouquillon:2003} for an extension 
to the polar motion), after dissipation of its rotational energy. There should remain free oscillations with negligible amplitude 
around the equilibrium, in the following we assume them as null, since they cannot be detected except for the Moon 
(Rambaux \& Williams 2011 \cite{Rambaux:2011}). It can be shown that, for rigid dynamics, between 2 and 4 Cassini States exist. In the context of natural satellites 
of the giant planets where the nodal precession rate is small with respect to the orbital frequency, the 4 Cassini States exist, and they induce an obliquity close to $k\frac{\pi}{2}$, $k$ being an integer (see Ward \& Hamilton 2004 \cite{Ward:2004} or Noyelles 2010 \cite{Noyelles:2010a}, Appendix B). The Cassini State 1, corresponding to $k=0$, i.e. a small obliquity, is a priori the most probable one, because it is stable and the primordial obliquity of the satellite is thought to be small.

\par In order to numerically simulate the rotational dynamics of the satellite, we need initial conditions that are actually very 
close to the equilibrium, that is perturbed by the orbital dynamics of the satellite. For that, we use the algorithm NAFFO 
(Noyelles et al. 2011 \cite{Noyelles:2011}), consisting in:

\begin{enumerate}

 \item A first numerical integration of the equations of the system, with initial conditions conveniently chosen,

 \item Frequency analysis of the solution and identification of the contributions depending on the free modes,

 \item Evaluation of the free modes at the time origin of the numerical simulation, and removal from the initial conditions,

\end{enumerate}
then the process is iterated until convergence. In a Hamiltonian framework as is the case here, Noyelles et al. 
\cite{Noyelles:2011} have shown that the convergence is quadratic in the amplitude of the free modes, provided that the 
quasi-periodic decomposition is exact, i.e. that the signal is indeed quasi-periodic, and that the numerical error has negligible 
impact. The proof is based on the d'Alembert characteristic (see e.g. Henrard 1974 \cite{Henrard:1974}), that gives a relation 
between the amplitudes $A_n$ and the arguments $\nu_nt$ in Eq.\ref{eq:naffc}. This algorithm has already been successfully applied 
in problem of rotational dynamics (Dufey et al. 2009 \cite{Dufey:2009}, Noyelles 2009 \cite{Noyelles:2009}, Robutel et al. 2011 
\cite{Robutel:2011}), in dynamics of exoplanetary systems (Couetdic et al. 2010 \cite{Couetdic:2010}), and in the analysis of 
ground-track resonances around Vesta (Delsate 2011 \cite{Delsate:2011}).

\subsection{The numerical tests}

\par The numerical algorithm we have just described has been used in different cases, dependent on the free parameters $\epsilon_3$, 
$\epsilon_4$ ($\approx$ polar flattening and equatorial ellipticity of the core), and $\delta=C_c/C$, representing the size of the 
core through its inertial polar momentum. In all our simulations we considered a kind of pseudo-Io, i.e. a satellite with physical
and dynamical properties close to the ones of the Galilean satellite of Jupiter Io, except that its orbit has constant eccentricity
and inclination. The numerical integrations are performed with the Adams-Bashforth-Moulton 10th order predictor–corrector integrator, with a tolerance of $10^{-14}$, and a step size of $5\times10^{-5}$ y $\approx1.8\times10^{-3}$ d.

\begin{table}[tbp]
 \centering
\caption{Physical and dynamical parameters ruling our pseudo-Io. We used the same as \cite{Henrard:2008}. The orbital frequency $n$
and the regression rate of the ascending orbital node $\dot{\ascnode}$ are taken from L1.2 ephemerides \cite{Lainey:2006}. The 
phases (initial conditions of the orbital angles) are arbitrarily chosen.\label{tab:pseudoIo}}
\begin{tabular}{c|c}
Parameter & Value \\
\hline
$GM_p$ (planet) & $1.261648547674763616\times10^{23}$ $km^3.s^{-2}$ \\
$GM$ (satellite) & $5955.5$ $km^3.s^{-2}$ \\
$R_p$ & $71492$ km \\
$J_{2p}$ & $1.4736\times10^{-2}$ \\
$J_2$ & $1.828\times10^{-3}$ \\
$C_{22}$ & $5.537\times10^{-4}$ \\
$C/(MR^2)$ & $0.376856$ \\
$a$ & $422029.958$ km \\
$e$ & $4.15\times10^{-3}$ \\
$I$ & $2.16$ arcmin \\
$n$ & $1297.2044725279755$ rad/y \\
$\dot{\varpi}$ & $0.97311853791375$ rad/y \\
$\dot{\ascnode}$ & $-0.8455888497945$ rad/y \\
$\lambda_o(0)$ & 0 \\
$\varpi_o(0)$ & $2$ rad \\
$\ascnode_o(0)$ & $0.1$ rad \\
$\epsilon_1=J_2\frac{MR^2}{C}$ & $4.85066\times10^{-3}$ \\
$\epsilon_2=2C_{22}\frac{MR^2}{C}$ & $2.93852\times10^{-3}$ \\
\hline
 \end{tabular}
\end{table}

\par We considered as reference values for the internal structure parameters $\delta=0.5$, $\epsilon_3=\epsilon_1$ and 
$\epsilon_4=\epsilon_2$, and we tested different pseudo-Ios with different values of these parameters.

\section{``Classical'' behavior}

\par We expect to have, at the Cassini State 1:

\begin{itemize}
 \item $\sigma=p-\lambda_o+\pi$ close to 0 because of the 1:1 spin-orbit resonance,
 \item $P$ close to $1$ (the norm of the angular momentum being close to nC),
 \item $\rho=\ascnode_o-h=\ascnode_o+r$ (third Cassini Law),
 \item $R$ close to $0$ (the obliquity being small),
 \item $J$ and $J_c$ close to $0$ (small polar motions of the satellite and its core),
\end{itemize}
the ``classical'' behavior being small oscillations around this equilibrium. We use it to define our first initial conditions, before
refining them with NAFFO.

  \subsection{In-depth study of a reference case}

\par We here present an in-depth study of a ``reference case'', considering $\epsilon_3=\epsilon_1$, $\epsilon_4=\epsilon_2$, 
and $\delta=0.5$. This study consists of a numerical estimation of the frequencies of the proper librations (Tab.\ref{tab:freqprop4e}),
and of a numerical decomposition of the canonical variables (Tab.\ref{tab:Pcas4e} to \ref{tab:etaxi2cas4e}).

\begin{table}[tbp]
\centering
\caption{Proper frequencies of the small oscillations around the equilibrium for $\epsilon_3=\epsilon_1$, 
$\epsilon_2=\epsilon_4$ and $\delta=0.5$. $n$ is the orbital frequency given in Tab.\ref{tab:pseudoIo}.\label{tab:freqprop4e}}
\begin{tabular}{r|rrr}
 & Frequency & Period & $\omega/n$\\
 & (rad/y) & (d) & \\
\hline
$\omega_u$ &  $243.4050908$ &   $9.42845$ & $0.187638$ \\
$\omega_v$ &    $4.1898509$ & $547.73630$ & $3.2299\times10^{-3}$ \\
$\omega_w$ &   $19.5319416$ & $117.49643$ & $0.015057$ \\
$\omega_z$ & $1334.4264821$ &   $1.71979$ & $1.028694$ \\
\hline
\end{tabular}
\end{table}

\par We recall that the orbital frequency $n$ is $1297.20447137$ rad/y (Lainey et al. 2006 \cite{Lainey:2006}). A comparison with 
Henrard \cite{Henrard:2008} lacks of significance since the physical model was different (the gravitational torque of Jupiter acting 
only on the mantle, while it acts on the whole satellite here), but we can see that, like Henrard, we find a proper frequency of the 
core $\omega_z$ close to the spin frequency of Io, that is also its orbital frequency since our satellite is locked in the 1:1 
spin-orbit resonance.

\par The Tab.\ref{tab:Pcas4e} to \ref{tab:etaxi2cas4e} give a quasi-periodic decomposition of the canonical variables with 
identification of the forced oscillations, i.e. the mean longitude of our pseudo-Io $\lambda_o$, the motion of its pericenter 
$\varpi_o$, and the motion of its orbital ascending node $\ascnode_o$. The phases are indicated at the time origin and allow to 
determine the presence of $\pi$ or $\pi/2$ in the identification.

\par Since our rotational model takes 4 degrees of freedom into
account, we can split the canonical variables and moments into 3 groups related to these degrees of freedom.

\par The first group $(\sigma, P)$ (Tab.\ref{tab:Pcas4e} \&  \ref{tab:sigmacas4e}) can be linked to the longitudinal motion. We can 
see that the mean position is the theoretical equilibrium $(\sigma=0,P=1)$ related to the 1:1 spin-orbit resonance, and there are small
oscillations around this equilibrium, related to the mean anomaly $\lambda_o-\varpi_o$ and its harmonics. We can see from the 
Tab.\ref{tab:sigmacas4e} that the deviation from the theoretical equilibrium does not exceed 2 arcmin for an eccentricity of 
$4.15\times10^{-3}$. This amplitude is proportional to the eccentricity (at least for small eccentricities, see e.g. Comstock \& Bills
2003 \cite{Comstock:2003}), that induces periodic variations of the planet-satellite distance.

\begin{table}[tbp]
\centering
\caption{The variable P-1. The series are in cosine.\label{tab:Pcas4e}}
\begin{tabular}{r|rrrrr}
 & Amplitude & Frequency & Phase & T   & Identification\\
 &           & (rad/y)   & (t=0) & (d) & \\
\hline
1 & $1.5156914\times10^{-4}$  & $1296.2313540$ &   $65.408^{\circ}$ & $1.77047$ & $\lambda_o-\varpi_o+\pi$ \\
2 & $6.6760683\times10^{-7}$  & $2592.4627080$ &  $-49.183^{\circ}$ & $0.88523$ & $2\lambda_o-2\varpi_o+\pi$ \\
3 & $3.8653845\times10^{-9}$  & $3888.6940620$ & $-163.775^{\circ}$ & $0.59016$ & $3\lambda_o-3\varpi_o+\pi$ \\
4 & $1.2702319\times10^{-9}$  &            $0$ &     $-180^{\circ}$ &  $\infty$ & cst \\
5 & $2.2957822\times10^{-11}$ & $5184.9254160$ &   $81.634^{\circ}$ & $0.44262$ & $4\lambda_o-4\varpi_o+\pi$ \\
6 & $1.5952860\times10^{-11}$ & $2596.1001228$ &  $-11.459^{\circ}$ & $0.88399$ & $2\lambda_o-2\ascnode$ \\
\hline
\end{tabular}
\end{table}

\begin{table}[tbp]
\centering
\caption{The resonant argument $\sigma$. The series are in cosine.\label{tab:sigmacas4e}}
\begin{tabular}{r|rrrrr}
 & Amplitude & Frequency & Phase & T   & Identification\\
 &           & (rad/y)   & (t=0) & (d) & \\
\hline
1 & $62.574$ arcsec            & $1296.2313540$ &  $-24.592^{\circ}$ & $1.77047$ & $\lambda_o-\varpi_o+\pi/2$ \\
2 &  $0.138$ arcsec            & $2592.4627080$ & $-139.183^{\circ}$ & $0.88523$ & $2\lambda_o-2\varpi_o+\pi/2$ \\
3 &  $0.053$ arcsec            & $3888.6940620$ &  $106.225^{\circ}$ & $0.59016$ & $3\lambda_o-3\varpi_o+\pi/2$ \\
4 & $2.03\times10^{-5}$ arcsec & $2596.1001228$ & $-101.459^{\circ}$ & $0.88399$ & $2\lambda_o-2\ascnode_o-\pi/2$ \\
5 & $2.37\times10^{-6}$ arcsec & $5184.9254164$ &   $-8.366^{\circ}$ & $0.44262$ & $4\lambda_o-4\varpi_o+\pi/2$ \\
6 & $1.21\times10^{-6}$ arcsec & $1299.8687682$ &  $166.868^{\circ}$ & $1.76551$ & $\lambda_o+\varpi_o-2\ascnode_o-\pi/2$ \\
\hline
\end{tabular}
\end{table}

\par The second group $(\rho,R)$ (Tab.\ref{tab:Rcas4e} \& \ref{tab:rhocas4e}) locates the angular momentum of the whole body with 
respect to the orbital plane. Once more, the angle can be averaged to $0$ with a instantaneous departure that does not exceed 2 
arcmin, this equilibrium is a consequence of the third Cassini law. It is also known that the mean obliquity, that can be derived 
from the mean value of $R$, is due to the interior structure and the regression rate of the orbital node (see e.g. Ward \& Hamilton
2004 \cite{Ward:2004}). We can also see that the oscillations are dominated by the mode $2\lambda_o-2\ascnode_o$, emphasizing an 
influence of the orbital node on this degree of freedom.

\begin{table}[tbp]
\centering
\caption{The variable R. The series are in cosine.\label{tab:Rcas4e}}
\begin{tabular}{r|rrrrr}
 & Amplitude & Frequency & Phase & T   & Identification\\
 &           & (rad/y)   & (t=0) & (d) & \\
\hline
1 & $2.5966515\times10^{-7}$  &            $0$ &        $0^{\circ}$ &  $\infty$ & cst \\
2 & $1.6920424\times10^{-10}$ & $2596.1001228$ &  $-11.459^{\circ}$ & $0.88399$ & $2\lambda_o-2\ascnode_o$ \\
3 & $3.1006440\times10^{-11}$ & $1296.2313540$ &   $65.408^{\circ}$ & $1.77047$ & $\lambda_o-\varpi_o+\pi$ \\
4 & $1.2914058\times10^{-12}$ & $3892.3314767$ &  $-36.051^{\circ}$ & $0.58960$ & $3\lambda_o-\varpi_o-2\ascnode_o+\pi/2$ \\
5 & $5.5914871\times10^{-13}$ &    $3.6374149$ & $-142.276^{\circ}$ & $630.924$ & $2\varpi_o-2\ascnode_o$ \\
\hline
\end{tabular}
\end{table}

\begin{table}[tbp]
\centering
\caption{The variable $\rho$. The series are in cosine.\label{tab:rhocas4e}}
\begin{tabular}{r|rrrrr}
 & Amplitude & Frequency & Phase & T   & Identification\\ 
 &           & (rad/y)   & (t=0) & (d) & \\
\hline
1 & $67.204$ arcsec & $2596.1001228$ &  $78.541^{\circ}$ & $0.88399$ & $2\lambda_o-2\ascnode_o+\pi/2$ \\
2 &  $1.542$ arcsec & $1296.2313540$ & $155.408^{\circ}$ & $1.77047$ & $\lambda_o-\varpi_o-\pi/2$ \\
3 &  $0.517$ arcsec & $3892.3314767$ &  $53.949^{\circ}$ & $0.58960$ & $3\lambda_o-\varpi_o-2\ascnode_o+\pi$ \\
4 &  $0.222$ arcsec &    $3.6374146$ & $-52.276^{\circ}$ & $630.924$ & $2\varpi_o-2\ascnode_o-3\pi/2$ \\
\hline
\end{tabular}
\end{table}

\par The third group involves the last two degrees of freedom $(\xi_1,\eta_1)$ (Tab.\ref{tab:etaxi1cas4e}) and $(\xi_2,\eta_2)$
(Tab.\ref{tab:etaxi2cas4e}), that are strongly coupled as shown by Henrard \cite{Henrard:2008}). They represent respectively the 
polar motion of the whole body and the orientation of the velocity field of the fluid. They are ruled by two kinds of small 
oscillations: fast ones due to harmonics of the proper mode $\lambda_o-\ascnode_o$, and slow ones due to the argument of the
pericenter $\varpi_o-\ascnode_o$.

\begin{table}[tbp]
\centering
\caption{The variable $\eta_1+\imath\xi_1$. The series are in complex exponential.\label{tab:etaxi1cas4e}}
\begin{tabular}{r|rrrrr}
 & Amplitude & Frequency & Phase & T   & Identification\\
 &           & (rad/y)   & (t=0) & (d) & \\
\hline
1 & $5.22646\times10^{-5}$ & $-1298.0500614$ & $-174.270^{\circ}$ &  $1.76799$ & $-\lambda_o+\ascnode_o-\pi$ \\
2 & $5.54231\times10^{-7}$ &  $1298.0500614$ &  $174.270^{\circ}$ &  $1.76799$ & $\lambda_o-\ascnode_o+\pi$ \\
3 & $2.81640\times10^{-7}$ &    $-1.8187074$ &   $71.138^{\circ}$ & $1261.849$ & $\ascnode_o-\varpi_o+\pi$ \\
4 & $2.20597\times10^{-7}$ &     $1.8187074$ &  $-71.138^{\circ}$ & $1261.849$ & $\varpi_o-\ascnode_o-\pi$ \\
5 & $5.96421\times10^{-9}$ & $-2594.2814154$ &  $-59.679^{\circ}$ &  $0.88461$ & $\varpi_o+\ascnode_o-2\lambda_o-\pi$ \\
6 & $3.66084\times10^{-9}$ &  $2594.2814154$ &   $59.679^{\circ}$ &  $0.88461$ & $2\lambda_o-\varpi_o-\ascnode_o+\pi$ \\
\hline
\end{tabular}
\end{table}

\begin{table}[tbp]
\centering
\caption{The variable $\eta_2+\imath\xi_2$. The series are in complex exponential.\label{tab:etaxi2cas4e}}
\begin{tabular}{r|rrrrr}
 & Amplitude & Frequency & Phase & T   & Identification\\
 &           & (rad/y)   & (t=0) & (d) & \\
\hline
1 & $7.36054\times10^{-5}$ &  $1298.0500614$ &   $-5.730^{\circ}$ & $1.76799$ & $\lambda_o-\ascnode_o$ \\
2 & $3.33184\times10^{-7}$ & $-1298.0500614$ &    $5.730^{\circ}$ & $1.76799$ & $-\lambda_o+\ascnode_o$ \\
3 & $2.03442\times10^{-7}$ &     $1.8187074$ &  $108.862^{\circ}$ & $1261.849$ & $\varpi_o-\ascnode_o$ \\
4 & $1.55904\times10^{-7}$ &    $-1.8187074$ & $-108.862^{\circ}$ & $1261.849$ & $\ascnode_o-\varpi_o$ \\
\hline
\end{tabular}
\end{table}

  \subsection{Influence of the parameters}

\par To characterise the influence of the internal structure parameters (i.e. $\epsilon_3$, $\epsilon_4$ and $\delta$), we quantify 
their effects on our outputs. We choose here to consider in particular the proper frequencies $\omega_u$ to $\omega_z$, and the mean 
value of $R$ (Tab.\ref{tab:infludeltae} to \ref{tab:influeps4e}).

\begin{table}[tbp]
\centering
\caption{Influence of the size of the core $\delta$, with $\epsilon_3=\epsilon_1$ and $\epsilon_4=\epsilon_2$.\label{tab:infludeltae}}
\begin{tabular}{r|rrrrr}
$\delta$ & $T_u$ (d) & $T_v$ (d) & $T_w$ (d) & $T_z$ (d) & $R^*$ \\
\hline
$0.1$ & $12.650$ & $453.259$ & $208.551$ & $1.745$ & $2.304\times10^{-7}$ \\
$0.2$ & $11.926$ & $480.369$ & $185.790$ & $1.741$ & $2.365\times10^{-7}$ \\
$0.3$ & $11.156$ & $504.669$ & $163.028$ & $1.736$ & $2.450\times10^{-7}$ \\
$0.4$ & $10.328$ & $526.965$ & $140.264$ & $1.729$ & $2.525\times10^{-7}$ \\
$0.5$ &  $9.428$ & $547.734$ & $117.496$ & $1.720$ & $2.597\times10^{-7}$ \\
$0.6$ &  $8.433$ & $567.278$ &  $94.723$ & $1.706$ & $2.680\times10^{-7}$ \\
$0.7$ &  $7.303$ & $585.780$ &  $71.939$ & $1.685$ & $2.760\times10^{-7}$ \\
$0.8$ &  $5.963$ & $603.294$ &  $49.130$ & $1.645$ & $2.842\times10^{-7}$ \\
$0.9$ &  $4.216$ & $619.394$ &  $26.230$ & $1.540$ & $2.922\times10^{-7}$ \\
\hline
\end{tabular}
\end{table}

\par We can see that all these outputs depend on the size of the core $\delta$ (Tab.\ref{tab:infludeltae}). In particular, the period 
of the free longitudinal librations $T_u$ follows the classical law (see e.g. Goldreich \& Peale 1966 \cite{Goldreich:1966}):

\begin{equation}
\label{eq:Tu}
 T_u\approx\frac{\pi}{n}\sqrt{\frac{C_m/\left(MR^2\right)}{3C_{22}}},
\end{equation}
yielding $T_u \propto \sqrt{1-\delta}$. We note that this period depends on the size of the core, while Henrard did not find any 
dependency in applying the gravitational torque just on the mantle. To check the influence of the shape of the core, we now present
the outputs with varying $\epsilon_3$ (Tab.\ref{tab:influeps3e}) and $\epsilon_4$ (Tab.\ref{tab:influeps4e}).

\begin{table}[tbp]
\centering
\caption{Influence of the polar flattening of the core $\epsilon_3$, with $\delta=0.5$ and $\epsilon_4=\epsilon_2$.
\label{tab:influeps3e}}
\begin{tabular}{r|rrrrr}
$\epsilon_3/\epsilon_1$ & $T_u$ (d) & $T_v$ (d) & $T_w$ (d) & $T_z$ (d) & $R^*$ \\
\hline
$0.2$ & $9.428$ & $6414.819$ & $117.118$ & $1.728$ & $1.040\times10^{-6}$ \\
$0.3$ & $9.428$ & $2491.673$ & $117.112$ & $1.727$ & $4.612\times10^{-7}$ \\
$0.4$ & $9.428$ & $1572.784$ & $117.121$ & $1.726$ & $3.592\times10^{-7}$ \\
$0.5$ & $9.428$ & $1163.452$ & $117.147$ & $1.725$ & $3.177\times10^{-7}$ \\
  $1$ & $9.428$ &  $547.734$ & $117.496$ & $1.720$ & $2.597\times10^{-7}$ \\
  $3$ & $9.428$ &  $254.876$ & $122.879$ & $1.695$ & $2.326\times10^{-7}$ \\
  $5$ & $9.428$ &  $210.742$ & $136.657$ & $1.668$ & $2.277\times10^{-7}$ \\ 
  $6$ & $9.428$ &  $200.875$ & $148.926$ & $1.655$ & $2.268\times10^{-7}$ \\
  $7$ & $9.428$ &  $194.454$ & $168.492$ & $1.641$ & $2.260\times10^{-7}$ \\
  $8$ & $9.428$ &  $189.284$ & $201.639$ & $1.628$ & $2.256\times10^{-7}$ \\
  $9$ & $9.428$ &  $185.621$ & $278.943$ & $1.616$ & $2.251\times10^{-7}$ \\
\hline
\end{tabular}
\end{table}

\par We recall that for Mercury, i.e. in the case of the 3:2 spin-orbit resonance, the flattening of the core $\epsilon_3$ alters 
the frequencies $\omega_v$ and $\omega_z$, but not the others ones. Here, the variations of the period of the free longitudinal
librations $T_u$ have only negligible variations, while the 3 other proper frequencies are affected. As for Mercury, the periods
$T_v$ and $T_z$ increase with $\epsilon_3$ getting closer to $0$, $T_z$ getting closer to the spin period $1.769$ d, and $T_v$ tending
to infinity. We also have an increase of the free wobble period when $\epsilon_3$ increases. We can note that it seems to be possible
to fine-tune the parameters ($\epsilon_3\approx7.7\epsilon_1$) to have a resonance between the free wobble and the free oscillations 
of the obliquity ($T_v=T_w$), but this is only anecdotal. This very peculiar case would require strict fine-tuning between the flattening of the body and of the core to occur, so we can consider it as very unlikely. Finally, the equilibrium position of the angular momentum, i.e. $R^*$, is 
shifted from the origin (here the normal to the orbit) when the core tends to be spherical (small $\epsilon_3$).

\par In the case of the 3:2 spin-orbit resonance, no significant influence of the equatorial ellipticity of the core had been 
detected. We here (Tab.\ref{tab:influeps4e}) see a small influence on $T_v$, $T_w$, $T_z$ and $R^*$, but that does not seem to be 
significant. Once more, the longitudinal librations seem not to be affected.

\begin{table}[tbp]
\centering
\caption{Influence of the  equatorial ellipticity of the core $\epsilon_4$, with $\delta=0.5$ and $\epsilon_3=\epsilon_1$.
\label{tab:influeps4e}}
\begin{tabular}{r|rrrrr}
$\epsilon_4/\epsilon_2$ & $T_u$ (d) & $T_v$ (d) & $T_w$ (d) & $T_z$ (d) & $R^*$ \\
\hline
  $0$ & $9.428$ &  $545.949$ & $117.771$ & $1.7199$ & $2.5996\times10^{-7}$ \\
$0.1$ & $9.428$ &  $546.128$ & $117.718$ & $1.7199$ & $2.5998\times10^{-7}$ \\
$0.5$ & $9.428$ &  $546.841$ & $117.563$ & $1.7199$ & $2.5954\times10^{-7}$ \\
  $1$ & $9.428$ &  $547.734$ & $117.496$ & $1.7198$ & $2.5967\times10^{-7}$ \\
  $3$ & $9.428$ &  $551.316$ & $118.652$ & $1.7195$ & $2.6070\times10^{-7}$ \\
  $5$ & $9.428$ &  $554.914$ & $122.283$ & $1.7193$ & $2.6069\times10^{-7}$ \\
 $10$ & $9.428$ &  $564.010$ & $149.248$ & $1.7186$ & $2.6248\times10^{-7}$ \\
\hline
\end{tabular}
\end{table}

\par As a reminder, Henrard \cite{Henrard:2005} found free periods of respectively $T_u=13.25$, $T_v=159.39$ and $T_w=229.85$ 
days in considering a rigid Io. The rigid value of $13.25$ days can be obtained in setting $C_m=C$ in Eq.\ref{eq:Tu}.

\section{Analysis of a bifurcation}

\par In the previous section, we do not present the behavior of the system for some physically possible values of the core shape
parameters $\epsilon_3$ and $\epsilon_4$. The reason is that for some range of these parameters, the system presents more complex
behaviors, that we here introduce. In particular, we assume since the beginning a ``classical'' Cassini State 1 in which null amplitudes
of the polar motions of the body $J$ and of the core $J_c$ define a stable equilibrium. In fact, this has not been checked yet, and
our numerical investigations have revealed that this equilibrium is unstable for instance for $\delta=0.5$, $\epsilon_3=10\epsilon_1$ 
and $\epsilon_4=0$.

  \subsection{Numerical characterisation of the equilibria}

\par
\begin{figure}[tbp]
 \centering
\begin{tabular}{cc}
 \includegraphics[width=5.6cm,height=3.3cm]{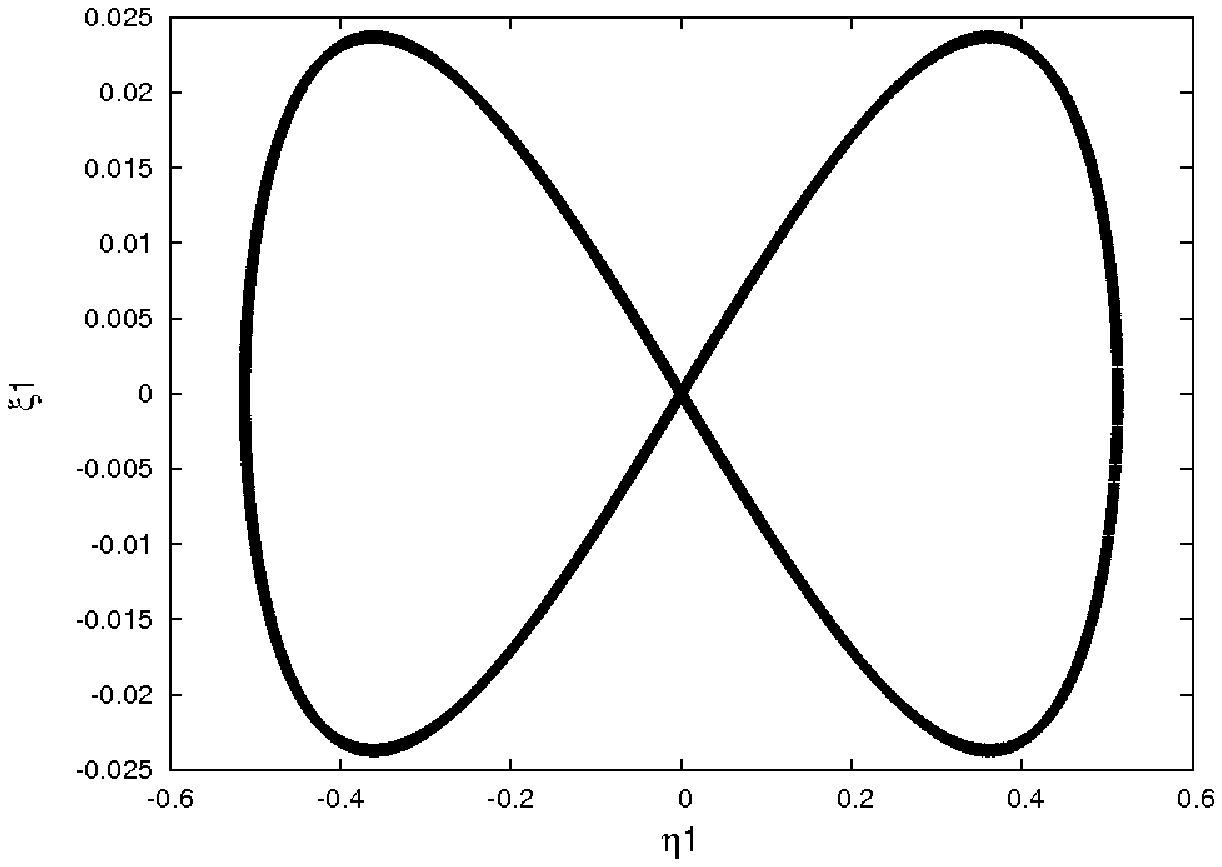} & \includegraphics[width=5.6cm,height=3.3cm]{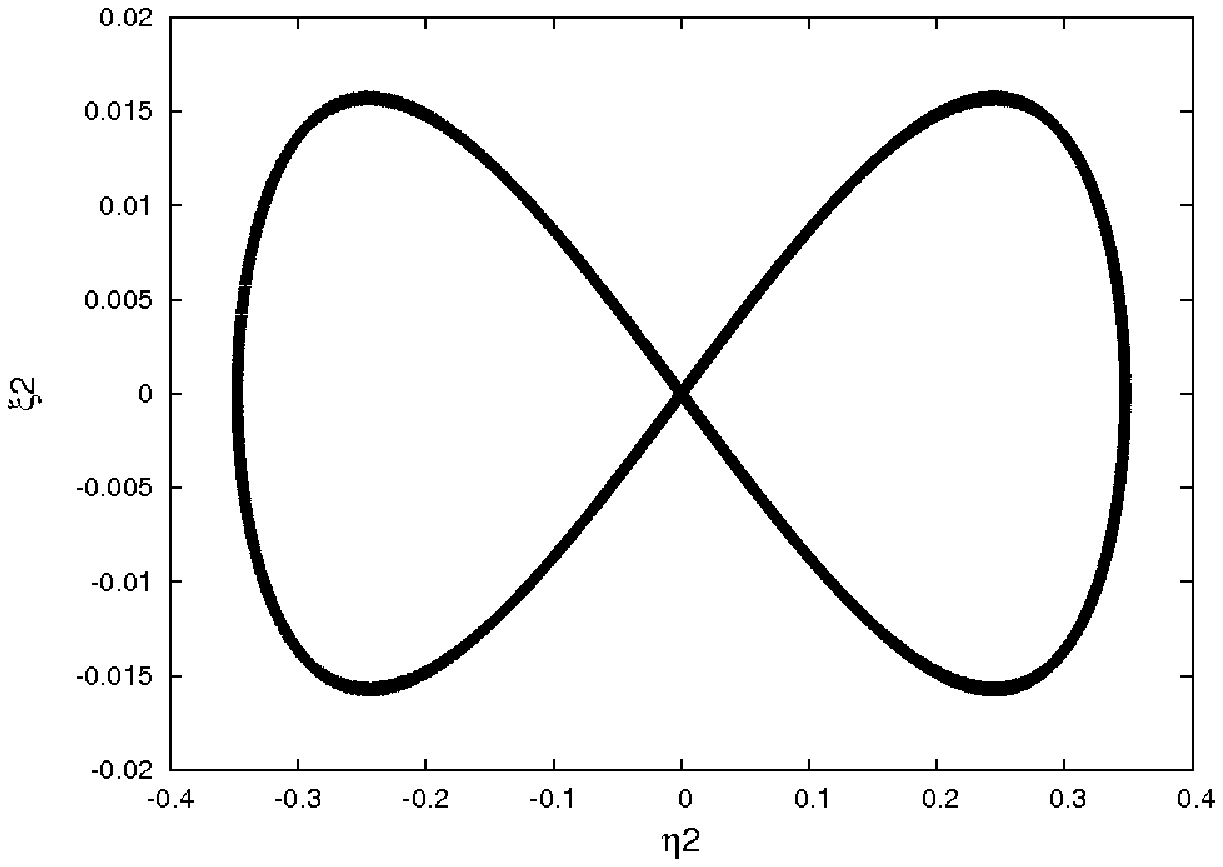}
\end{tabular}
\caption{Trajectory passing close to the equilibrium $\xi_1=\xi_2=\eta_1=\eta_2$ for $\delta=0.5$, $\epsilon_3=10\epsilon_1$ and $\epsilon_4=0$. The left panel shows the polar motion of the whole body, and the right one is related to the pseudo-core. We can see that the trajectory does not librate around this equilibrium, but presents a butterfly-shape, that suggests the presence of 2 new stable equilibria.\label{fig:zzunstable}}
\end{figure}

 A simulation of the behavior of the system for $\delta=0.5$, $\epsilon_3=10\epsilon_1$ and $\epsilon_4=0$ gives a butterfly shape
for the outputs related to the polar motion of the body $(\eta_1,\xi_1)$ and of the velocity field of the fluid $(\eta_2,\xi_2)$ for
the solution passing close to the equilibrium defined by $J=J_c=0$ (Fig.\ref{fig:zzunstable}). It suggests that this equilibrium is in 
fact unstable, and that two new stable equilibria appear.

These equilibria have been reached thanks to NAFFO. The quasi-periodic decompositions of the solution corresponding to the equilibrium
$(\xi_1=\xi_2=0,\eta_1\approx0.25,\eta_2\approx-0.17)$ are given in Tab.\ref{tab:Pcas26e} to \ref{tab:etaxi2cas26e}. The other 
equilibrium is symmetrical to this one, i.e. corresponds to $(\xi_1=\xi_2=0,\eta_1\approx-0.25,\eta_2\approx0.17)$.

\begin{table}[tbp]
\centering
\caption{The variable P-1 for $\epsilon_3=10\epsilon_1$ and $\epsilon_4=0$. The series are in cosine.\label{tab:Pcas26e}}
\begin{tabular}{r|rrrrr}
 & Amplitude & Frequency & Phase & T   & Identification\\
 &           & (rad/y)   & (t=0) & (d) & \\
\hline
1 & $2.81032\times10^{-3}$  & $0$            & $0$                & $\infty$    & cst \\
2 & $1.56939\times10^{-4}$  & $1296.2313540$ &   $65.408^{\circ}$ &  $1.77047$ & $\lambda_o-\varpi_o+\pi$ \\
3 & $6.77223\times10^{-7}$  & $2592.4627080$ &  $-49.183^{\circ}$ &  $0.88523$ & $2\lambda_o-2\varpi_o+\pi$ \\
4 & $5.20072\times10^{-8}$  & $1298.0500614$ &  $174.270^{\circ}$ &  $1.76799$ & $\lambda_o-\ascnode_o+\pi$ \\
5 & $3.80792\times10^{-8}$  &    $1.8187074$ &  $288.862^{\circ}$ & $1261.849$ & $\varpi_o-\ascnode_o+\pi$ \\
6 & $3.92165\times10^{-9}$  & $3888.6940620$ & $-163.775^{\circ}$ &  $0.59016$ & $3\lambda_o-3\varpi_o+\pi$ \\
7 & $3.50511\times10^{-10}$ & $2594.2814154$ &   $59.679^{\circ}$ &  $0.88461$ & $2\lambda_o-\varpi_o-\ascnode_o+\pi$ \\
\hline
\end{tabular}
\end{table}

\begin{table}[tbp]
\centering
\caption{The resonant argument $\sigma$ for $\epsilon_3=10\epsilon_1$ and $\epsilon_4=0$. The series are in cosine.\label{tab:sigmacas26e}}
\begin{tabular}{r|rrrrr}
 & Amplitude & Frequency & Phase & T   & Identification\\
 &           & (rad/y)   & (t=0) & (d) & \\
\hline
1 & $63.973$ arcsec & $1296.2313540$ &  $-24.592^{\circ}$ &  $1.77047$ & $\lambda_o-\varpi_o+\pi/2$ \\
2 &  $0.139$ arcsec & $2592.4627080$ & $-139.183^{\circ}$ &  $0.88523$ & $2\lambda_o-2\varpi_o+\pi/2$ \\
3 &  $0.113$ arcsec & $1298.0500614$ &   $84.270^{\circ}$ &  $1.76799$ & $\lambda_o-\ascnode_o+\pi/2$ \\
4 & $5.4\times10^{-4}$ arcsec & $3888.6940620$ &  $106.225^{\circ}$ & $0.59016$ & $3\lambda_o-3\varpi_o+\pi/2$ \\
5 & $3.0\times10^{-5}$ arcsec & $2594.2814154$ &  $149.679^{\circ}$ & $0.88461$ & $2\lambda_o-\varpi_o-\ascnode_o+3\pi/2$ \\
6 & $2.2\times10^{-5}$ arcsec & $1294.4126466$ &   $46.546^{\circ}$ & $1.77295$ & $\lambda_o-2\varpi_o+\ascnode_o+3\pi/2$ \\
7 & $1.6\times10^{-5}$ arcsec & $2596.1001228$ & $-101.459^{\circ}$ & $0.88399$ & $2\lambda_o-2\ascnode_o-\pi/2$ \\
\hline
\end{tabular}
\end{table}

\begin{table}[tbp]
\centering
\caption{The variable R for $\epsilon_3=10\epsilon_1$ and $\epsilon_4=0$. The series are in cosine.\label{tab:Rcas26e}}
\begin{tabular}{r|rrrrr}
 & Amplitude & Frequency & Phase & T   & Identification\\
 &           & (rad/y)   & (t=0) & (d) & \\
\hline
1 & $3.3172485\times10^{-7}$  & $0$            &        $0^{\circ}$ & $\infty$   & cst \\
2 & $3.0899861\times10^{-7}$  &    $1.8187074$ &  $108.862^{\circ}$ & $1261.849$ & $\varpi_o-\ascnode_o$ \\
3 & $1.4728954\times10^{-10}$ & $2596.1001228$ &  $-11.459^{\circ}$ &  $0.88399$ & $2\lambda_o-2\ascnode_o$ \\
4 & $7.9302319\times10^{-11}$ & $2592.4627080$ &  $-49.183^{\circ}$ &  $0.87986$ & $2\lambda_o-2\varpi_o+\pi$ \\
5 & $2.0015617\times10^{-11}$ & $1296.2313540$ &   $65.408^{\circ}$ &  $1.77047$ & $\lambda_o-\varpi_o+\pi$ \\
6 & $1.5218917\times10^{-11}$ & $2594.2814154$ & $-120.321^{\circ}$ &  $0.88461$ & $2\lambda_o-\varpi_o-\ascnode_o$ \\
7 & $8.1094453\times10^{-12}$ & $1294.4126466$ &   $46.546^{\circ}$ &  $1.77295$ & $\lambda_o-2\varpi_o+\ascnode_o+3\pi/2$ \\
8 & $4.4761150\times10^{-12}$ & $1298.0500614$ &  $174.270^{\circ}$ &  $1.76799$ & $\lambda_o-\ascnode_o+\pi$ \\
\hline
\end{tabular}
\end{table}

\begin{table}[tbp]
\centering
\caption{The variable $\rho$ for $\epsilon_3=10\epsilon_1$ and $\epsilon_4=0$. The series are in cosine.\label{tab:rhocas26e}}
\begin{tabular}{r|rrrrr}
 & Amplitude & Frequency & Phase & T   & Identification\\
 &           & (rad/y)   & (t=0) & (d) & \\
\hline
 1 & $39.163^{\circ}$ &    $1.8187074$ &   $18.862^{\circ}$ & $1261.849$ & $\varpi_o-\ascnode_o-\pi/2$ \\
 2 & $13.384^{\circ}$ &    $3.6374148$ &  $127.724^{\circ}$ &  $630.924$ & $2\varpi_o-2\ascnode_o-\pi/2$ \\
 3 &  $6.099^{\circ}$ &    $5.4561222$ & $-123.414^{\circ}$ &  $420.616$ & $3\varpi_o-3\ascnode_o-\pi/2$ \\
 4 &  $3.127^{\circ}$ &    $7.2748300$ &  $-14.552^{\circ}$ &  $315.462$ & $4\varpi_o-4\ascnode_o-\pi/2$ \\
 5 &  $1.710^{\circ}$ &    $9.0935369$ &   $94.310^{\circ}$ &  $252.370$ & $5\varpi_o-5\ascnode_o-\pi/2$ \\
 6 &  $58.431$ arcmin &   $10.9122443$ & $-156.828^{\circ}$ &  $210.308$ & $6\varpi_o-6\ascnode_o-\pi/2$ \\
 7 &  $34.233$ arcmin &   $12.7309517$ &  $-47.966^{\circ}$ &  $180.264$ & $7\varpi_o-7\ascnode_o-\pi/2$ \\
 8 &  $20.474$ arcmin &   $14.5496590$ &   $60.896^{\circ}$ &  $157.731$ & $8\varpi_o-8\ascnode_o-\pi/2$ \\
 9 &  $12.440$ arcmin &   $16.3683661$ &  $169.758^{\circ}$ &  $140.205$ & $9\varpi_o-9\ascnode_o-\pi/2$ \\
10 &   $7.653$ arcmin &   $18.1870733$ &  $-81.380^{\circ}$ &  $126.185$ & $10\varpi_o-10\ascnode_o-\pi/2$ \\
11 &   $4.755$ arcmin &   $20.0057804$ &   $27.482^{\circ}$ &  $114.714$ & $11\varpi_o-11\ascnode_o-\pi/2$ \\
12 &   $2.979$ arcmin &   $21.8244872$ &  $136.344^{\circ}$ &  $105.154$ & $12\varpi_o-12\ascnode_o-\pi/2$ \\
13 &   $1.880$ arcmin &   $23.6431936$ & $-114.794^{\circ}$ &   $97.065$ & $13\varpi_o-13\ascnode_o-\pi/2$ \\
14 &   $1.715$ arcmin & $2596.1001226$ &   $78.541^{\circ}$ &  $0.88399$ & $2\lambda_o-2\ascnode_o+\pi/2$ \\
15 &   $1.193$ arcmin &   $25.4618992$ &   $-5.932^{\circ}$ &   $90.132$ & $14\varpi_o-14\ascnode_o-\pi/2$ \\
16 &   $1.168$ arcmin & $2597.9188300$ & $-172.597^{\circ}$ &  $0.88337$ & $2\lambda_o+\varpi_o-3\ascnode_o-3\pi/2$ \\
17 &  $52.836$ arcsec & $2594.2814149$ &  $-30.321^{\circ}$ &  $0.88461$ & $2\lambda_o-\varpi_o-\ascnode_o-3\pi/2$ \\
18 &  $47.694$ arcsec & $2599.7375373$ & $-116.265^{\circ}$ &  $0.88276$ & $2\lambda_o+2\varpi_o-4\ascnode_o-3\pi/2$ \\
19 &  $45.673$ arcsec &   $27.2806037$ &  $102.930^{\circ}$ &   $84.123$ & $15\varpi_o-15\ascnode_o-\pi/2$ \\
20 &  $32.468$ arcsec & $2601.5562447$ &   $45.127^{\circ}$ &  $0.88214$ & $2\lambda_o+3\varpi_o-5\ascnode_o-3\pi/2$ \\
21 &  $29.268$ arcsec &   $29.0993068$ & $-148.208^{\circ}$ &   $78.866$ & $16\varpi_o-16\ascnode_o-\pi/2$ \\
22 &  $22.103$ arcsec & $2603.3749520$ &  $153.989^{\circ}$ &  $0.88152$ & $2\lambda_o+4\varpi_o-6\ascnode_o-3\pi/2$ \\
23 &  $18.829$ arcsec &   $30.9180086$ &  $-39.346^{\circ}$ &   $74.226$ & $17\varpi_o-17\ascnode_o-\pi/2$ \\
24 &  $15.046$ arcsec & $2605.1936596$ &  $-97.149^{\circ}$ &  $0.88091$ & $2\lambda_o+5\varpi_o-7\ascnode_o-3\pi/2$ \\
\hline
\end{tabular}
\end{table}

\begin{table}[tbp]
\centering
\caption{The variable $\eta_1+\imath\xi_1$ for $\epsilon_3=10\epsilon_1$. The series are in complex exponential.
\label{tab:etaxi1cas26e}}
\begin{tabular}{r|rrrrr}
 & Amplitude & Frequency & Phase & T   & Identification\\
 &           & (rad/y)   & (t=0) & (d) & \\
\hline
 1 & $0.2501568$ & $0$ & $0^{\circ}$ & $\infty$ & cst \\
 2 & $3.36571\times10^{-5}$ & $-1296.2313540$ & $-155.408^{\circ}$ &  $1.77047$ & $\varpi_o-\lambda_o-3\pi/2$ \\
 3 & $7.44538\times10^{-6}$ &  $1296.2313540$ &  $155.408^{\circ}$ &  $1.77047$ & $\lambda_o-\varpi_o+3\pi/2$ \\
 4 & $5.19058\times10^{-6}$ & $-1298.0500614$ & $-174.270^{\circ}$ &  $1.76799$ & $-\lambda_o+\ascnode_o-\pi$ \\
 5 & $1.72624\times10^{-6}$ &     $1.8187074$ &  $-71.138^{\circ}$ & $1261.849$ & $\varpi_o-\ascnode_o+\pi$ \\
 6 & $1.69988\times10^{-6}$ &    $-1.8187074$ &   $71.138^{\circ}$ & $1261.849$ & $\ascnode_o-\varpi_o-\pi$ \\
 7 & $5.90500\times10^{-7}$ &  $1298.0500614$ &  $174.270^{\circ}$ &  $1.76799$ & $\lambda_o-\ascnode_o+\pi$ \\
 8 & $4.69722\times10^{-8}$ &  $2592.4627080$ &  $-49.183^{\circ}$ &  $0.88523$ & $2\lambda_o-2\varpi_o+\pi$ \\
 9 & $3.95648\times10^{-8}$ & $-2592.4627080$ &   $43.183^{\circ}$ &  $0.88523$ & $2\varpi_o-2\lambda_o-\pi$ \\
10 & $3.31650\times10^{-9}$ &  $2594.2814161$ &   $59.679^{\circ}$ &  $0.88461$ & $2\lambda_o-\varpi_o-\ascnode_o+\pi$ \\
\hline
\end{tabular}
\end{table}

\begin{table}[tbp]
\centering
\caption{The variable $\eta_2+\imath\xi_2$ for $\epsilon_3=10\epsilon_1$. The series are in complex exponential.
\label{tab:etaxi2cas26e}}
\begin{tabular}{r|rrrrr}
 & Amplitude & Frequency & Phase & T   & Identification\\
 &           & (rad/y)   & (t=0) & (d) & \\
\hline
 1 & $0.1690165$            & $0$             &      $180^{\circ}$ & $\infty$   & cst \\
 2 & $3.25350\times10^{-5}$ &  $1296.2313540$ & $-114.592^{\circ}$ &  $1.77047$ & $\lambda_o-\varpi_o$ \\
 3 & $7.01986\times10^{-6}$ &  $1298.0500614$ &   $-5.730^{\circ}$ &  $1.76799$ & $\lambda_o-\ascnode_o$ \\
 4 & $1.16880\times10^{-6}$ &    $-1.8187074$ & $-108.862^{\circ}$ & $1261.848$ & $\ascnode_o-\varpi_o$ \\
 5 & $1.15291\times10^{-6}$ &     $1.8187074$ &  $108.862^{\circ}$ & $1261.848$ & $\varpi_o-\ascnode_o$ \\
 6 & $1.09791\times10^{-6}$ & $-1296.2313540$ &  $-65.408^{\circ}$ &  $1.77047$ & $\varpi_o-\lambda_o-\pi$ \\
 7 & $2.37221\times10^{-7}$ & $-1298.0500614$ &    $5.730^{\circ}$ &  $1.76799$ & $\ascnode_o-\lambda_o$ \\
 8 & $8.29840\times10^{-9}$ &  $2592.4627079$ &  $130.817^{\circ}$ &  $0.88523$ & $2\lambda_o-2\varpi_o$ \\
 9 & $4.26308\times10^{-9}$ &  $2594.2814150$ &   $59.679^{\circ}$ &  $0.88461$ & $2\lambda_o-\varpi_o-\ascnode_o+\pi$ \\
10 & $1.54598\times10^{-9}$ & $-2592.4627077$ & $-130.817^{\circ}$ &  $0.88523$ & $2\varpi_o-2\lambda_o$ \\
\hline
\end{tabular}
\end{table}

\par We can see from these tables that the difference is not only in $(\xi_1,\eta_1,\xi_2,\eta_2)$. The difference for the degree of
freedom related to the longitudinal behavior $(\sigma,P)$ is striking. First, we can see a significant departure ($2.81\times10^{-3}$)
from the expected mean P, i.e. 1 (Tab.\ref{tab:Pcas26e}). We also note significant longitudinal librations related to the combination
of proper modes $\lambda_o-\ascnode_o$ (Tab.\ref{tab:sigmacas26e}), that did not appear in the ``classical'' behavior 
(Tab.\ref{tab:sigmacas4e}).

\par The difference is even more important for the degree of freedom related to the location of the angular momentum, i.e. $(\rho,R)$
(Tab.\ref{tab:rhocas26e} \& \ref{tab:Rcas26e}). In this case, we can see large oscillations associated with the argument of the 
pericenter $\varpi_o-\ascnode_o$. It is known that a motion due to the position of the pericenter has the eccentricity as physical
cause, while our eccentricity is only $4.15\times10^{-3}$, the peak-to-peak oscillations of $\rho$ reaching $80^{\circ}$. So, we can 
expect higher oscillations for bigger eccentricities.

\par In this case, the shift of P led us to change iteratively the value of the constant $P_c$ so that it remains equal to $\delta<P>$.
We have seen that a change of $P_c$ yields a significant difference on the locations of the stable equilibria, that is the reason why
the mean values of $\eta_1+\imath\xi_1$ and $\eta_2+\imath\xi_2$ we give in Tab.\ref{tab:etaxi1cas26e},\ref{tab:etaxi2cas26e} are 
significantly different from the ones that can be guessed from Fig.\ref{fig:zzunstable}.

  \subsection{Analytical study}

\par In order to understand the appearance of 2 new stable equilibria, we propose a simplified analytical study of the problem. This
study consists in starting from the Hamiltonian $\mathcal{H}$ (Eq.\ref{equ:hamiltout}), in expressing the oscillating angle (respectively 
$\sigma=p-\lambda_o+\pi$ because of the 1:1 spin-orbit resonance, and $\rho=\ascnode_o-h$ because of the third Cassini Law), in 
averaging over the circulating ones, to deduce a secular Hamiltonian yielding the equilibria. All these calculations have been 
performed thanks to Maple software.

\par The starting point is the Hamiltonian $\mathcal{H}$ (Eq.\ref{equ:hamiltout}) in which the coordinates of the perturber (i.e. a pseudo-Jupiter 
if we consider a pseudo-Io) $x$ and $y$ are replaced thanks to Eq.\ref{equ:passage} with

\begin{eqnarray}
x_i&=&-\left(\cos\ascnode_o\cos(\lambda_o-\ascnode_o)-\cos I_o\sin\ascnode_o\sin(\lambda_o-\ascnode_o)\right), \\
y_i&=&-\left(\sin\ascnode_o\cos(\lambda_o-\ascnode_o)+\cos I_o\cos\ascnode_o\sin(\lambda_o-\ascnode_o)\right), \\
z_i&=&-\sin I_o\sin(\lambda_o-\ascnode_o).
\end{eqnarray}
We here neglect the influence of the eccentricity. \\
Then the following canonical transformation is performed

\begin{equation}\label{eq:chres}
\begin{array}{lll}
\sigma=p-\lambda_o+\pi, & \hspace{2cm} & P, \\
\rho=\ascnode_o+r, & \hspace{2cm} & R, \\
\xi_1, & \hspace{2cm} & \eta_1, \\
\xi_2, & \hspace{2cm} & \eta_2. \\
\end{array} \\
\end{equation}
Since this transformation, involving $\lambda_o$ and $\ascnode_o$, is time-dependent, we must add $-nP+\dot{\ascnode}R$ to the 
Hamiltonian. $\sigma$ and $\rho$ are oscillating arguments that can be averaged to $0$, while $\lambda_o$ and $\ascnode_o$ are 
circulating. 

\par A first-order averaging of the Hamiltonian is performed, then the Hamilton equations are derived, i.e.

\begin{equation}\label{eq:hequations}
\begin{array}{lll}
\frac{d\sigma}{dt}=\frac{\partial \mathcal{H}}{\partial P}, & \hspace{2cm} & 
\frac{dP}{dt}=-\frac{\partial \mathcal{H}}{\partial \sigma}, \\
& \\
\frac{d\rho}{dt}=\frac{\partial \mathcal{H}}{\partial R}, & \hspace{2cm} & \frac{dR}{dt}=-\frac{\partial \mathcal{H}}{\partial r}, \\
& \\
\frac{d\xi_1}{dt}=\frac{\partial \mathcal{H}}{\partial \eta_1}, & \hspace{2cm} & 
\frac{d\eta_1}{dt}=-\frac{\partial \mathcal{H}}{\partial \xi_1}, \\
& \\
\frac{d\xi_2}{dt}=\frac{\partial \mathcal{H}}{\partial \eta_2}, & \hspace{2cm} & 
\frac{d\eta_2}{dt}=-\frac{\partial \mathcal{H}}{\partial \xi_2}, \\
\end{array} \\
\end{equation}
the equilibria corresponding to null time derivatives of the variables and associated moments, i.e. the right-hand side of these equations vanish. 
The numerical exploration drove us to neglect the influence of the inclination and the obliquity ($I=0$, $R=0$), and to consider $\xi_1$ and $\xi_2$ as null at the equilibrium. These approximations allowed us to simplify the system, and we finally find with a good agreement the equilibrium values of $P$, $\eta_1$ and $\eta_2$ in solving numerically the following equations:

\begin{eqnarray}
 \frac{1}{n}\frac{d\sigma}{dt} & =& -1+\frac{P-P_c}{1-\delta} +\frac{\eta_1^2}{2(1-\delta)^2}
\left(\epsilon_1-\epsilon_2-\delta\epsilon_3+\delta\epsilon_4\right)+\frac{\eta_2^2}{2(1-\delta)} \label{eq:ndpdt} \\
& +& \frac{\eta_1\eta_2\left(P_c-\eta_2^2/4\right)}{2\left(1-\delta\right)\sqrt{PP_c-P\eta_2^2/4-P_c\eta_1^2/4+\eta_1^2\eta_2^2/16}}
\left(1+\frac{\epsilon_1-\epsilon_2-\delta\epsilon_3+\delta\epsilon_4}{1-\delta}\right),\nonumber
\end{eqnarray}

\begin{eqnarray}
\frac{1}{n}\frac{d\xi_1}{dt} & = & \frac{\eta_1P}{(1-\delta)^2}\left(\epsilon_1-\epsilon_2-\delta\epsilon_3+\delta\epsilon_4\right)+
\frac{\eta_1P_c}{1-\delta}+\frac{\eta_1^3}{2(1-\delta)^2}\left(-\epsilon_1+\epsilon_2+\delta\epsilon_3-\delta\epsilon_4\right)-
\frac{\eta_1\eta_2^2}{2(1-\delta)} \nonumber \\
& + & \frac{\eta_1^2\eta_2\left(\eta_2^2/4-P_c\right)}{4(1-\delta)\sqrt{PP_c-P\eta_2^2/4-P_c\eta_1^2/4+\eta_1^2\eta_2^2/16}}
\left(1+\frac{\epsilon_1-\epsilon_2-\delta\epsilon_3+\delta\epsilon_4}{1-\delta}\right) \label{eq:ndxi1dt} \\
& + & \frac{\eta_2}{1-\delta}\sqrt{PP_c-P\eta_2^2/4-P_c\eta_1^2/4+\eta_1^2\eta_2^2/16}
\left(1+\frac{\epsilon_1-\epsilon_2-\delta\epsilon_3+\delta\epsilon_4}{1-\delta}\right), \nonumber 
\end{eqnarray}
and
\begin{eqnarray}
\frac{1}{n}\frac{d\xi_2}{dt} & = & \frac{\eta_2P_c}{(1-\delta)^2}\left(\epsilon_1-\epsilon_2+\left(\frac{1}{\delta}-2\right)\epsilon_3+
\left(2-\frac{1}{\delta}\right)\epsilon_4\right)+\frac{\eta_2P}{1-\delta} \nonumber \\
&+&\frac{\eta_2^3}{2(1-\delta)^2}\left(-\epsilon_1+\epsilon_2+
\left(2-\frac{1}{\delta}\right)\epsilon_3+\left(\frac{1}{\delta}-2\right)\epsilon_4\right)-
\frac{\eta_1^2\eta_2}{2(1-\delta)} \label{eq:ndxi2dt} \\
& + & \frac{\eta_1\eta_2^2\left(\eta_1^2/4-P\right)}{4(1-\delta)\sqrt{PP_c-P\eta_2^2/4-P_c\eta_1^2/4+\eta_1^2\eta_2^2/16}}
\left(1+\frac{\epsilon_1-\epsilon_2-\delta\epsilon_3+\delta\epsilon_4}{1-\delta}\right) \nonumber \\
& + & \frac{\eta_1}{1-\delta}\sqrt{PP_c-P\eta_2^2/4-P_c\eta_1^2/4+\eta_1^2\eta_2^2/16}
\left(1+\frac{\epsilon_1-\epsilon_2-\delta\epsilon_3+\delta\epsilon_4}{1-\delta}\right)\nonumber.
\end{eqnarray}

\par For $\epsilon_3=10\epsilon_1$, $\epsilon_4=0$ and $\delta=0.5$, the real roots of this system are

\begin{itemize}
  \item $P=1.046772470$, $\eta_1=1.446908787$, $\eta_2=-1.023119016$
  \item $P=1.002812138$, $\eta_1=0.2502391659$, $\eta_2=-0.1690724173$
  \item $P=1$, $\eta_1=\eta_2=0$
  \item $P=1.002812138$, $\eta_1=-0.2502391659$, $\eta_2=0.1690724173$
  \item $P=0.3489241565$, $\eta_1=0.8353731582$, $\eta_2=-0.5606980247$
\end{itemize}
while they are, for $\epsilon_3=9\epsilon_1$, $\epsilon_4=0$ and $\delta=0.5$:

\begin{itemize}
  \item $P=1.041484268$, $\eta_1=-1.443249317$, $\eta_2=-1.020531379$
  \item $P=0.3471614224$, $\eta_1=-0.8332603709$, $\eta_2=-0.5892040580$
  \item $P=1$, $\eta_1=\eta_2=0$
  \item $P=0.9978852209$, $\eta_1=-2.010693353$, $\eta_2=-1.421011855$.
\end{itemize}
So, we can see for $P\approx1$ and $|\eta_1|$, $|\eta_2|<0.5$, an appearance of 2 additional equilibria. In order to test the validity of this analytical study, we propose (Tab.\ref{tab:compbifurk}) a short comparison between its results and the numerical results, in 3 cases where the 2 
equilibria appear. We can see a significant discrepancy for the first case, where $\epsilon_3=9.45\epsilon_1$ and $\epsilon_4=0$. In 
this case, the equilibria are close to the origin $\eta_1=\eta_2=0$, while a good agreement is reached for the other two cases,
where the equilibrium values of $\eta_1$ and $\eta_2$ are bigger. The observed discrepancy can be due to the neglect of the 
obliquity, the inclination and the eccentricity.

%\begin{table}[tbp]
% \centering
%\caption{Location of a new stable equilibrium, determined analytically (a) thanks to Eq.\ref{eq:ndpdt} to \ref{eq:ndxi2dt} and 
%numerically (n), for $\delta=0.5$. The last column, $A$, gives the amplitude of the 1.77-d longitudinal librations, obtained in our 
%numerical code. Here, only the equilibrium corresponding to $\eta_1>0$ and $\eta_2<0$ has been considered. In all these cases, 
%another stable equilibrium exists in changing the signs of $\eta_1$ and $\eta_2$.\label{tab:compbifurk}}
%  \begin{tabular}{cc|ccccccc}
%   $\epsilon_3/\epsilon_1$ & $\epsilon_4/\epsilon_2$ & $P-1$ (n) & $P-1$ (a) & $\eta_1$ (n) & $\eta_1$ (a) & 
%$\eta_2$ (n) & $\eta_2$ (a) & A \\
%\hline
%$9.45$ & $0$   & $1.60164\times10^{-4}$ & $4.06798\times10^{-4}$ & $0.06184$ & $0.09722$ & $-0.04182$ & $-0.06574$ & $0.027$ as \\
%  $10$ & $0$   & $2.81214\times10^{-3}$ & $2.81214\times10^{-3}$ & $0.25015$ & $0.25024$ & $-0.16902$ & $-0.16907$ & $0.113$ as \\
%  $10$ & $0.3$ & $1.95008\times10^{-3}$ & $1.95008\times10^{-3}$ & $0.20992$ & $0.20998$ & $-0.14872$ & $-0.14191$ & $0.093$ as \\
%\hline
%  \end{tabular}
%\end{table}

\begin{table}[tbp]
 \centering
\caption{Location of a new stable equilibrium, determined analytically (a) thanks to Eq.\ref{eq:ndpdt} to \ref{eq:ndxi2dt} and 
numerically (n), for $\delta=0.5$. The last column, $A$, gives the amplitude of the 1.77-d longitudinal librations, obtained in our 
numerical code. Here, only the equilibrium corresponding to $\eta_1>0$ and $\eta_2<0$ has been considered. In all these cases, 
another stable equilibrium exists in changing the signs of $\eta_1$ and $\eta_2$.\label{tab:compbifurk}}
  \begin{tabular}{cc|ccccccc}
   $\epsilon_3/\epsilon_1$ & $\epsilon_4/\epsilon_2$ & $P-1$ (n) & $P-1$ (a) & $\eta_1$ (n) & $\eta_1$ (a) & 
$\eta_2$ (n) & $\eta_2$ (a) \\
\hline
$9.45$ & $0$   & $1.5831\times10^{-4}$ & $4.0680\times10^{-4}$ & $0.0608$ & $0.0972$ & $-0.0411$ & $-0.0657$ \\
  $10$ & $0$   & $2.8103\times10^{-3}$ & $2.8121\times10^{-3}$ & $0.2502$ & $0.2502$ & $-0.1690$ & $-0.1691$ \\
  $10$ & $0.3$ & $1.9482\times10^{-3}$ & $1.9501\times10^{-3}$ & $0.2099$ & $0.2100$ & $-0.1418$ & $-0.1419$ \\
\hline
  \end{tabular}
\end{table}

\par We now propose to study the existence of these 2 additional equilibria. Since their existence is linked to the stability of the 
equilibrium corresponding to $\eta_i,\xi_i=0$, $P=1$ and $P_c=\delta$, we in fact study this stability. In setting $\xi_1=\xi_2=0$,
$P=1$ and $P_c=\delta$ in the averaged Hamiltonian, we get the quantity $\mathcal{S}$:

\begin{eqnarray}
  \mathcal{S}(\eta_1,\eta_2) & = & \alpha-1+\frac{1-\delta+\eta_1^2\delta+\eta_2^2-\eta_1^2\eta_2^2/2}{2(1-\delta)} \nonumber \\
& + & \epsilon_1\left(-\frac{3}{2}+\frac{\eta_1^2+\eta_2^2\delta-\left(\eta_1^4+\eta_2^4\right)/4}{2(1-\delta)^2}+
\frac{\alpha}{1-\delta}\right) \nonumber \\
& + & \epsilon_2\left(-\frac{3}{2}-\frac{\eta_1^2+\eta_2^2\delta-\left(\eta_1^4+\eta_2^4\right)/4}{2(1-\delta)^2}-
\frac{\alpha}{1-\delta}\right) \label{eq:surface} \\
& + & (\epsilon_3-\epsilon_4)\left(\frac{-\delta\eta_1^2+\eta_2^2(1-2\delta)+\delta\eta_1^4/4+
\eta_2^4/2(1-1/(2\delta))}{2(1-\delta)^2}-\frac{\delta\alpha}{1-\delta}\right) \nonumber
\end{eqnarray}
with

\begin{equation}
  \label{equ:alpha}
  \alpha=\frac{\eta_1\eta_2}{1-\delta}\sqrt{\delta-\frac{\eta_1^2\delta+\eta_2^2}{4}+\frac{\eta_1^2\eta_2^2}{16}}.
\end{equation}
We do not call $\mathcal{S}$ ``Hamiltonian'' since two variables, i.e. $\xi_1$ and $\xi_2$, are sets to constants, while their
associated momenta $\eta_1$ and $\eta_2$ vary. The study is now equivalent to the investigation of the extrema of the surface defined
by the Eq.\ref{eq:surface}. In fact we study the point defined by $\eta_1=\eta_2=0$, we know thanks to previous calculations that it 
gives null first-order derivatives of $\mathcal{S}$. The topological nature of this point can be investigated in studying the second
order partial derivatives of $\mathcal{S}$. We consider the Hessian matrix

\begin{eqnarray}
\mathcal{M}&=&\left(\begin{array}{cc}
\frac{\partial^2\mathcal{S}}{\partial\eta_1^2} &  \frac{\partial^2\mathcal{S}}{\partial\eta_1\partial\eta_2} \\
\frac{\partial^2\mathcal{S}}{\partial\eta_2\partial\eta_1} &  \frac{\partial^2\mathcal{S}}{\partial\eta_2^2}
\end{array}\right) \label{equ:matrixx} \\
&=&\frac{1}{(\delta-1)^2}\left(\begin{array}{cc}
\epsilon_1-\epsilon_2+\delta(\epsilon_4-\epsilon_3+1-\delta) & 
\sqrt{\delta}(1-\delta+\epsilon_1-\epsilon_2+\delta(\epsilon_4-\epsilon_3)) \\
\sqrt{\delta}(1-\delta+\epsilon_1-\epsilon_2+\delta(\epsilon_4-\epsilon_3)) & 
1-\delta+\delta(\epsilon_1-\epsilon_2-2\epsilon_3+2\epsilon_4)
\end{array}\right).\nonumber
\end{eqnarray}
A minimum (corresponding to a stable equilibrium) is reached when the two eigenvalues of the Hessian, $\lambda_{1,2}$, are positive.
We have:

\begin{eqnarray}
  \lambda_1&=&\beta+\frac{\sqrt{\Delta}}{2} \\
  \lambda_2&=&\beta-\frac{\sqrt{\Delta}}{2}
\end{eqnarray}
with

\begin{equation}
  \beta=\frac{1-\delta^2+(\epsilon_1-\epsilon_2)(1+\delta)+(\epsilon_3-\epsilon_4)(1-3\delta)}{2}
\end{equation}
and

\begin{eqnarray}
  \Delta & = & \left(1-\delta^2\right)^2-2(\epsilon_1-\epsilon_2)(1-7\delta+7\delta^2-\delta^3)
+2(\epsilon_3-\epsilon_4)(1-3\delta-\delta^2+3\delta^3) \nonumber \\
 & + & (\epsilon_1-\epsilon_2)^2(1+\delta)^2+2(\epsilon_1\epsilon_4+\epsilon_2\epsilon_3-\epsilon_1\epsilon_3
-\epsilon_2\epsilon_4)(1-2\delta+5\delta^2) \nonumber \\
 & + & (\epsilon_3-\epsilon_4)^2(1-2\delta+\delta^2+4\delta^3). \label{eq:gdelta}
\end{eqnarray}
Numerical evaluations show that $\lambda_1$ is always positive, and that $\lambda_2$ is usually positive, except for the interior
parameters given in Tab.\ref{tab:compbifurk}. In these peculiar cases, we have $\lambda_1\lambda_2<0$, so the considered point 
($\eta_1=\eta_2=0$) is a saddle point.

\par This study shows that the equilibrium corresponding to $J=J_c=0$ is unstable for $\lambda_2=0$. This condition is independent of
the mean motion and is applicable to any body in 1:1 spin-orbit resonance, in which the interior model of a rigid mantle, a fluid core
and a small solid inner core composed of dense material is realistic. We have also neglected the effect of the orbital inclination and of the regression of
the ascending node. This approximation is relevant, since most of the natural satellites of the giant planets have inclinations of the order of a few arcmin, and the nodal regression of Io is one of the
most rapid in the Solar System. Since here this approximation gives good results, it should be available for most of the Solar System
bodies in a comparable dynamical situation.

\subsection{Effect on the observable variables}

\par We now consider the influence of this peculiar behavior on the observable parameters, i.e. data that could be observed if our 
pseudo-Io were real and if it were observed with enough accuracy. In particular, they have to refer to the mantle since its rotation
is actually the rotation of the surface. These observable data can be deduced from the canonical variables, that give a complete 
mathematical description of the system.

\par A complete derivation of the observable outputs can be found in (Noyelles et al. \cite{Noyelles:2010}), we here choose to 
represent the following quantities:

\begin{itemize}

  \item the mean obliquity of the mantle $<K_m>$,

  \item the mean amplitude of the polar motion of the mantle $<J_m>$,

  \item the mean amplitude of the polar motion of the core $<J_c>$.

\end{itemize}

\par
\begin{figure}[ht]
  \centering
\begin{tabular}{cc}
  \includegraphics[width=3cm,height=5cm]{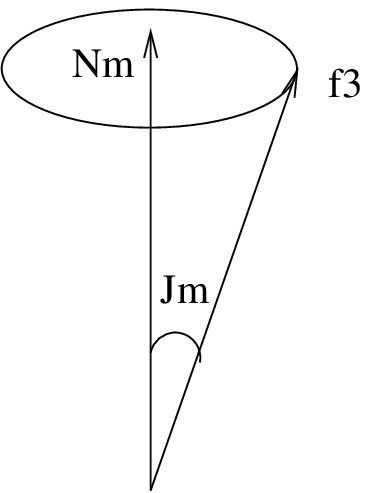} & \includegraphics[width=4cm,height=5cm]{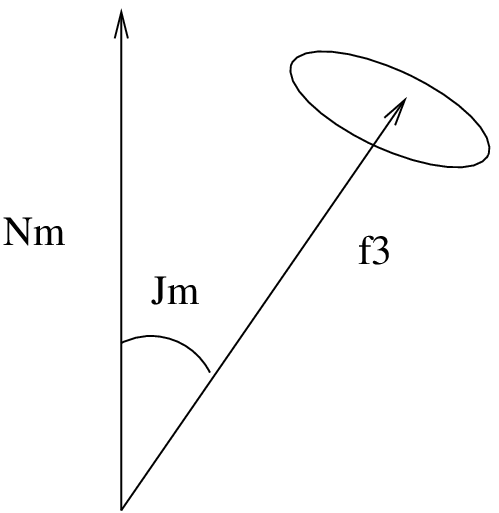} \\
  $\delta=0.5$, $\epsilon_3/\epsilon_1=\epsilon_4/\epsilon_2=1$ & $\delta=0.5$, $\epsilon_3/\epsilon_1=10$, $\epsilon_4=0$
\end{tabular}
 \caption{Location of the North Pole of the mantle of the body (located by $\vec{f_3}$) with respect to its angular momentum 
$\vec{N_m}$ in the classical case (left) and with a highly flattened core (right).\label{fig:Jm26}}
\end{figure}

 All these results are obtained thanks to frequency analysis, and they are gathered in Tab.\ref{tab:observa}. We can see
that the stable equilibria that appear induce a forcing of the polar motion of the surface (or mantle) of our pseudo Io 
(Fig.\ref{fig:Jm26}), that can reach $15^{\circ}$. In (Noyelles 2008 \cite{Noyelles:2008b}) we found a forcing of the polar motion 
of a rigid Titan, due to a resonance between the free wobble and the forced precession of Titan's perihelion. We considered it as a 
possible explanation for the super-synchronous rotation of Titan, before it was observed (Stiles et al. 2008 \& 2010 
\cite{Stiles:2008}). This is different here, since no resonance appears.

\begin{table}[tbc]
\centering
 \caption{The first case is the reference one, and the other ones correspond to the cases where two additional stable equilibria 
appear.\label{tab:observa}}
\begin{tabular}{cc|ccc}
 $\epsilon_3/\epsilon_1$ & $\epsilon_4/\epsilon_2$ & $<K_m>$ & $<J_m>$ & $<J_c>$ \\
\hline
$1$ & $1$ & $2.299$ am & $0.155$ as & $21.470$ as \\
$9.45$ & $0$ & $9.065$ am & $3.632^{\circ}$ & $3.335^{\circ}$ \\
$10$ & $0$ & $37.542$ am & $14.975^{\circ}$ & $13.726^{\circ}$ \\
$10$ & $0.3$ & $31.220$ am & $12.555^{\circ}$ & $11.516^{\circ}$ \\
\hline
\end{tabular}
\end{table}

\section{Orientation of the angular momentum}

\par Among the Third Cassini Law (see e.g. Cassini (1693) \cite{Cassini:1693} or Colombo (1966) \cite{Colombo:1966}), the equilibrium 
orientation of the total 
angular momentum of the body is assumed to be in the Cassini State 1. As a consequence, the angular momentum, the normal to the orbital
plane and the normal to the Laplace Plane are coplanar, the Laplace Plane being a reference plane based on the precessional motion
of the orbital ascending node, that minimizes the variations of the inclination of the considered body. There are in fact several ways
 to define this plane, as for instance in (Yseboodt et al. 2006 \cite{Yseboodt:2006}) or in (D'Hoedt et al. \cite{DHoedt:2009}). A 
difficulty is: how to consider a constant reference plane if the precession rate of the ascending node is not constant? Should we 
average over a ``long enough'' time interval, or over a time-interval suitable to the observations of a space mission?

\par The reader can find in (Noyelles 2009 \cite{Noyelles:2009}) a discussion on the choice of an ``appropriate'' reference plane 
depending on the variations of the orbital inclination, that allows the argument $\rho=\ascnode_o-h$ to librate. It is shown that, for the rotation of a 
rigid body in 1:1 spin-orbit resonance, if the satellite orbits close to its parent planet, the precessional motion is ruled by the
oblateness of the planet (its $J_2$) and so its precession rate is close to be constant. In such a case, choosing the equatorial plane 
of the planet as a reference plane to describe the behavior of the angular momentum of the body can be a convenient choice. However, 
when the satellite orbits far from its parent planet as it is the case for Titan or Callisto, the reference plane for the nodal 
precession is shifted because of the Solar gravitational perturbation. In such a case, considering the planet's equatorial plane as 
the reference plane could either result in a oscillating rotation node $h$ as it is the case for Titan (Noyelles et al. 2008 
\cite{Noyelles:2008}), either result in an erratic apparent behavior due to an improper choice of the reference plane, as is the case for 
Callisto (Noyelles 2009 \cite{Noyelles:2009}).

\par
\begin{figure}[tbp]
 \centering
\begin{tabular}{cc}
 \includegraphics[width=5.6cm,height=4cm]{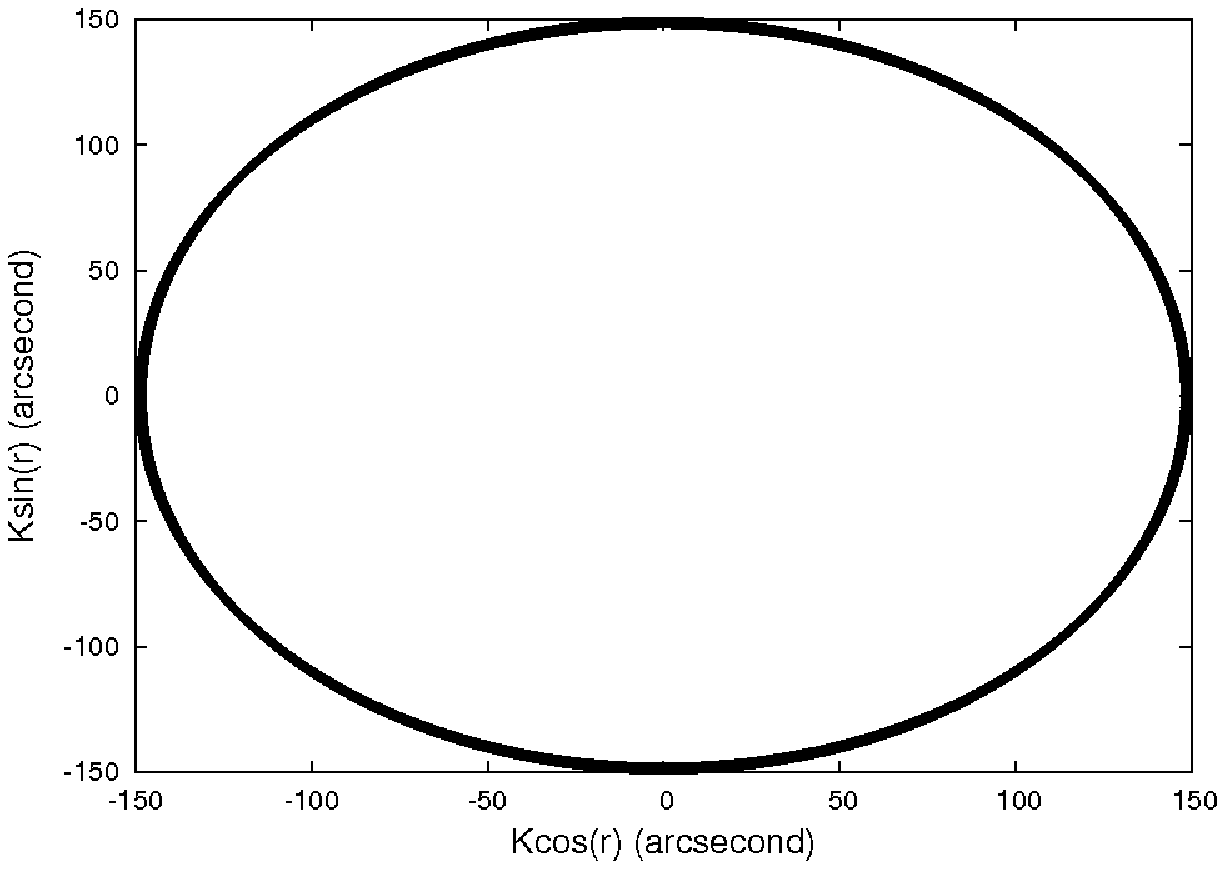} & \includegraphics[width=5.6cm,height=4cm]{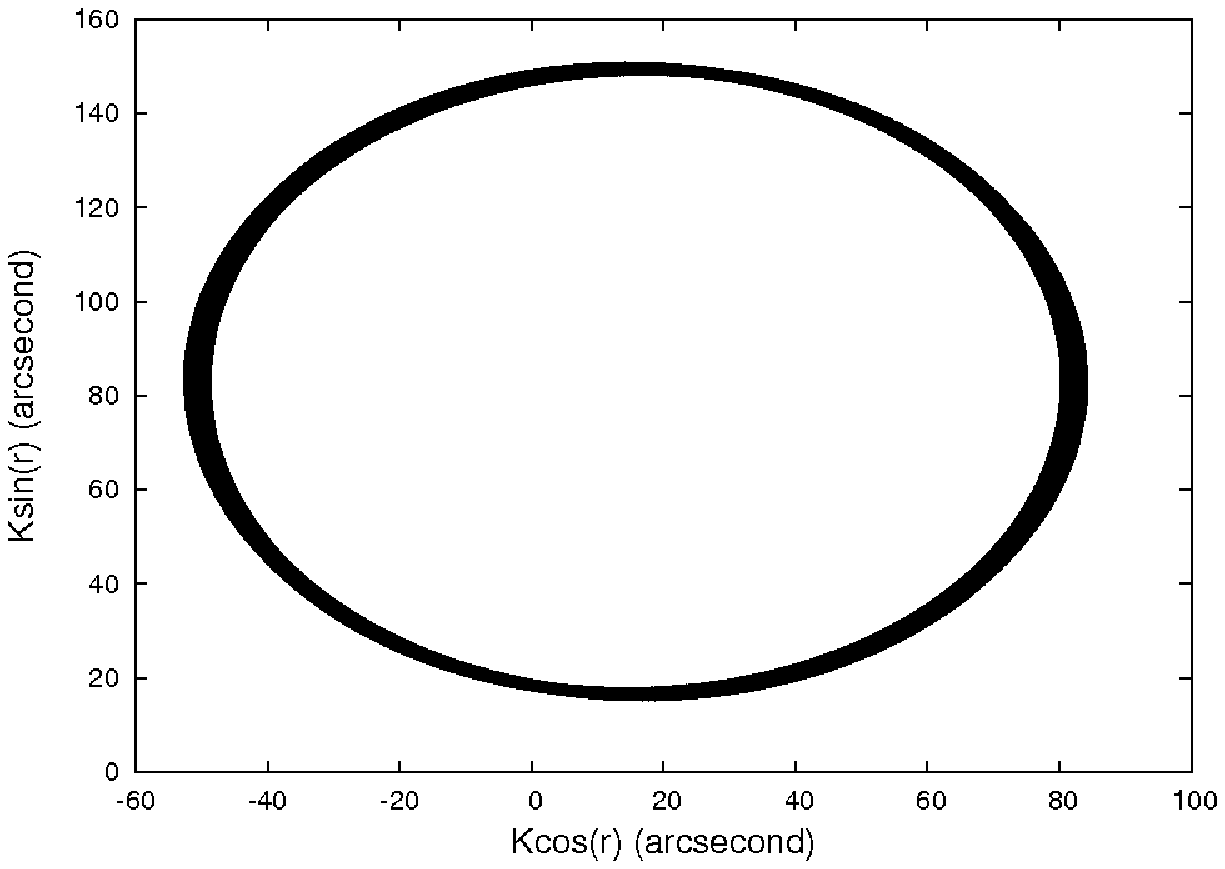} \\
$\delta=0.5$, $\epsilon_3=\epsilon_1$, $\epsilon_4=\epsilon_2$ & $\delta=0.5$, $\epsilon_3=0$, $\epsilon_4=\epsilon_2$
\end{tabular}
\caption{Behavior of the orientation of the angular momentum of our pseudo Io $K\exp\imath r$, with 2 different internal structure 
models, in the inertial reference frame. The right panel shows a shift of this motion that is not on averaged at the 
origin.\label{fig:Kr48}}
\end{figure}

 In our case of a pseudo-Io with a constant regression of the node, no ``strange'' behavior is expected. In particular, the 
Tab.\ref{tab:rhocas4e} supports the assumption of a quasi-periodic behavior of the difference of the nodes $\rho$. However, we have 
found a different behavior for a small flattening of the core $\epsilon_3$ (Fig.\ref{fig:Kr48} and Tab.\ref{tab:Ker8}) resulting in 
a significant shift of the mean equilibrium orientation of the total angular momentum. This shift seems to be not constant but a 
long-period oscillation, the period being $\approx57,000$ years. We call $\nu$ this oscillation.

\begin{table}[tbp]
\centering
\caption{The variable $K\exp\left(\imath r\right)$ for $\delta=0.5$, $\epsilon_3=0$ and $\epsilon_2=\epsilon_4$. The series are in 
complex exponential and the amplitudes in arcseconds. We can note a nearly constant component $\nu$, that has a negligible in the 
usual case.\label{tab:Ker8}}
\begin{tabular}{r|rrrrr}
 & Amplitude & Frequency & Phase & T   & Identification\\
 & (arcsec)  & (rad/y)   & (t=0) & (d) & \\
\hline
1 & $84.516$ & $1.093\times10^{-4}$ & $78.801^{\circ}$ & $2.1\times10^{7}$ & $\nu$ \\
2 & $66.484$ &     $0.8455888$ &   $-5.730^{\circ}$ & $2714.006$ & $-\ascnode_o$ \\
3 &  $0.022$ & $-2595.2545339$ &    $5.730^{\circ}$ &    $0.884$ & $\ascnode_o-2\lambda_o$ \\
4 &  $0.006$ &  $1296.2314632$ &  $-35.742^{\circ}$ &    $1.770$ & $\lambda_o-\varpi_o+\nu$ \\
5 &  $0.006$ & $-1296.2312447$ & $-166.558^{\circ}$ &    $1.770$ & $\varpi_o-\lambda_o+\nu$ \\
6 &  $0.005$ & $-1295.3857651$ &  $-71.138^{\circ}$ &    $1.772$ & $\varpi_o-\lambda_o-\ascnode_o-\pi$ \\
7 &  $0.005$ &  $1297.0769428$ &   $59.679^{\circ}$ &    $1.769$ & $\lambda_o-\varpi_o-\ascnode_o+\pi$ \\
\hline
\end{tabular}
\end{table}

\par In (Noyelles et al. 2010 \cite{Noyelles:2010}), we had found a particular behavior for small $\epsilon_3$, that we attributed to 
the exact resonance between the Free Core Nutation frequency $\omega_z$ and the spin frequency. We also noticed an asymptotic behavior of the free 
frequency $\omega_v$ that tended to $0$ (and the period $T_v$ to infinity) when $\epsilon_3$ tended to $0$. This last behavior is here
observed as well as can be seen in Tab.\ref{tab:influeps3e}. This is confirmed by some tests at $\epsilon_3=\epsilon_1/10$ suggesting
$T_v=9933.75$ days. However, even if the free period $T_z$ gets closer to the spin period of $1.76799$ day, it does not seem to 
reach it. So we cannot speak of resonant behavior, it seems more likely to be a kind of singularity at $\epsilon_3=0$.

\begin{figure}[tbp]
 \centering
\begin{tabular}{cc}
 \includegraphics[width=5.6cm,height=4cm]{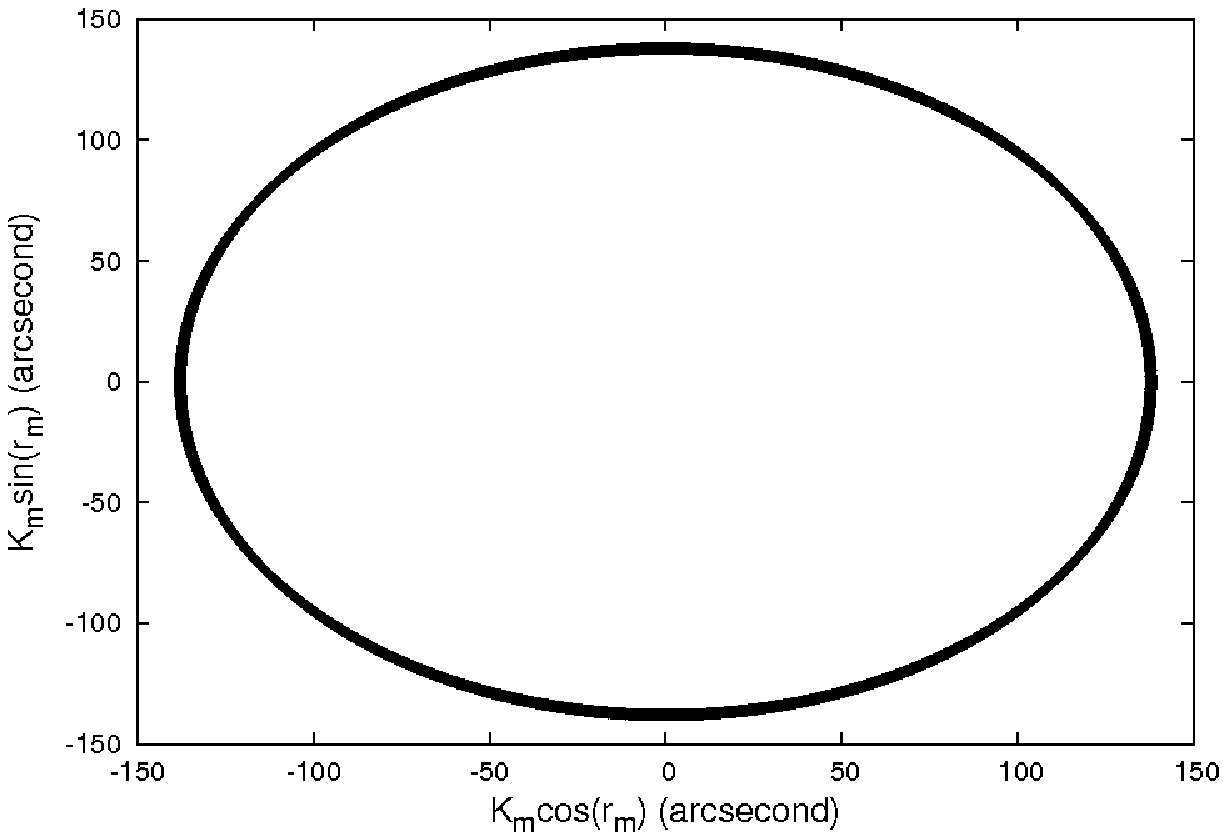} & \includegraphics[width=5.6cm,height=4cm]{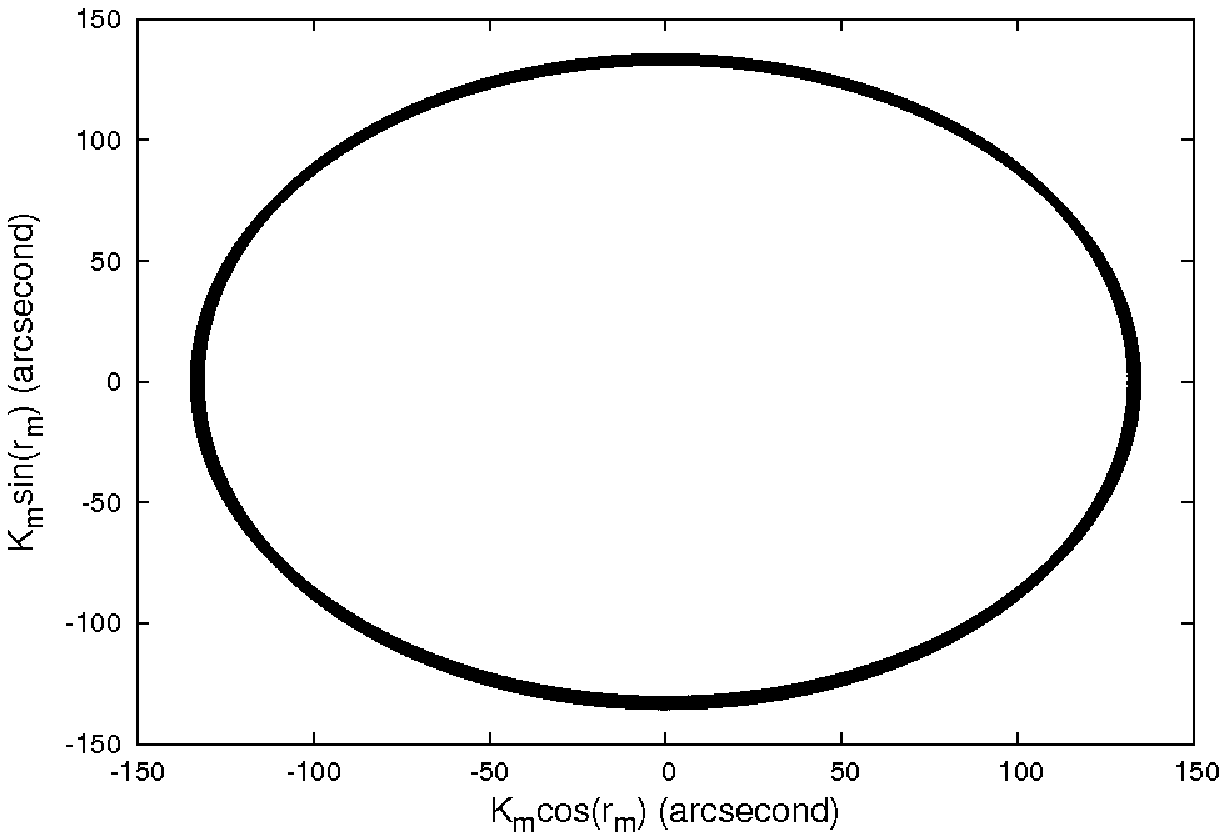} \\
$\delta=0.5$, $\epsilon_3=\epsilon_1$, $\epsilon_4=\epsilon_2$ & $\delta=0.5$, $\epsilon_3=0$, $\epsilon_4=\epsilon_2$
\end{tabular}
\caption{Behavior of the orientation of the angular momentum of the mantle (i.e. the surface) of our pseudo Io $K_m\exp\imath r_m$, 
with 2 different internal structure models, in the inertial reference frame. Contrary to the total angular momentum 
(Fig.\ref{fig:Kr48}), it does not exhibit particular behavior.\label{fig:Krm48}}
\end{figure}

\par The Fig.\ref{fig:Krm48} shows the orientation of the angular momentum of the mantle/surface, that does not exhibit this shift. So,
if such a situation would occur (i.e. very small polar flattening of the core), the equatorial/ring plane of the planet could be an
acceptable reference plane to describe the orientation of this axis. In fact, a physical signature of this dynamics remains in the 
core, we indeed get a mean $J_c$ of $\approx3$ arcmin for $\epsilon_3=0$ while we have $<J_c>\approx21$ arcsec for 
$\epsilon_3=\epsilon_1$.

\section{Conclusion}

\par In this study we have presented the behavior of a pseudo-Io orbit on a low eccentric orbit around its parent planet, with a 
uniform nodal regression and a constant inclination, in considering it as a two-layer body composed of a rigid mantle and a fluid 
triaxial core. This model can be applied to study the rotation of most differentiated natural satellites.

\par We have described the ``usual'' case, consisting of small oscillations around the expected equilibrium, i.e. synchronous rotation with 
a small obliquity and no polar motion, but we also have, especially for a highly flattened core, another behavior resulting in a polar motion
forced by several degrees. Another peculiar behavior is when the polar flattening of the core 
is very small. In this last case we have a forcing of the obliquity of the full body, but not of its mantle, so there should be no 
observational evidence of this phenomenon. From a mathematical point of view, this could be due to a kind of singularity in the 
parameter $\epsilon_3$. 

\par This study aimed at exploring the behavior of a model, its application to real bodies would require to consider complete 
ephemerides. This would add additional forcing frequencies complicating the dynamics of the system. New behavior cannot a priori be 
excluded.

\par A possibility to improve the model would be to consider nonlinear phenomena in the fluid, but this is another story\ldots

\begin{acknowledgements}
Numerical simulations were made on the local computing ressources (\emph{Cluster URBM-SYSDYN}) at the University of Namur 
(FUNDP, Belgium). The author is indebted to Nicolas Delsate and Julien Frouard for fruitful discussions. BN is F.R.S.-FNRS 
post-doctoral research fellow.
\end{acknowledgements}

\appendix

\section{The NAFF algorithm}

\par  The frequency analysis algorithm that we use is based on Laskar's original idea, named NAFF as Numerical Analysis of the 
Fundamental Frequencies (see for instance Laskar 1993 \cite{Laskar:1993} for the method, and Laskar 2005 \cite{Laskar:2005} for the 
convergence proofs). It aims at identifying the coefficients $a_k$ and $\omega_k$ of a complex signal $f(t)$ obtained numerically 
over a finite time span $[-T;T]$  and verifying

\begin{equation}
\label{equ:naff}
f(t) \approx \sum_{k=1}^na_k\exp(\imath\omega_kt),
\end{equation}
where $\omega_k$ are real frequencies and $a_k$ complex coefficients. If the signal $f(t)$ is real, its frequency spectrum is 
symmetric and the complex amplitudes associated with the frequencies $\omega_k$ and $-\omega_k$ are complex conjugates. The 
frequencies and amplitudes associated are found with an iterative scheme. To determine the first frequency $\omega_1$, one searches 
for the maximum of the amplitude of 

\begin{equation}
\label{equ:philas}
\phi(\omega)=<f(t),\exp(\imath\omega t)>,
\end{equation}
where the scalar product $<f(t),g(t)>$ is defined by

\begin{equation}
\label{equ:prodscal}
<f(t),g(t)>=\frac{1}{2T}\int_{-T}^T f(t)g(t)^*\chi(t) dt,
\end{equation}
$g(t)^*$ being the complex conjugate of $g(t)$. $\chi(t)$ is a weight function alike a Hann or a Hamming window, i.e. a 
positive function verifying

\begin{equation}
\label{equ:poids}
\frac{1}{2T}\int_{-T}^T \chi(t) dt=1.
\end{equation}
Using such a window can help the determination in reducing the amplitude of secondary minima in the transform (\ref{equ:prodscal}). 
Its use is optional.
\par Once the first periodic term $\exp(\imath\omega_1t)$ is found, its complex amplitude $a_1$ is obtained by orthogonal 
projection, and the process is started again on the remainder $f_1(t)=f(t)-a_1\exp(\imath\omega_1t)$. The algorithm stops when 
two detected frequencies are too close to each other, what alters their determinations, or when the number of detected terms reaches 
a limit set by the user. This algorithm is very efficient, except when two frequencies are too close to each other. In that case, the 
algorithm is not confident in its accuracy and stops. When the difference between two frequencies is larger than twice the frequency 
associated with the length of the total time interval, the determination of each fundamental frequency is not perturbed by the other 
ones. Although the iterative method suggested by Champenois \cite{Champenois:1998} allows to reduce this distance, some troubles may 
remain. In our particular case, these problems are likely to arise because of the proximity between the free frequency of the core 
$\omega_z$ and the frequency of the spin.


\end{document}